\documentclass[12pt]{amsart}
\usepackage[normalem]{ulem}

\advance\textheight\topskip
\usepackage{amsmath,amsfonts,amsthm,amssymb,amscd}
\usepackage[mathscr]{eucal}
\allowdisplaybreaks
\usepackage{hyperref}
\usepackage{times}
\usepackage[normalem]{ulem}

\usepackage[curve,matrix,arrow]{xy}

\numberwithin{equation}{section}
\makeatletter
\@addtoreset{equation}{section}
\@addtoreset{subsubsection}{section}

\def\@secnumfont{\bfseries}
\def\subsubsection{\@startsection{subsubsection}{3}%
  \z@{.5\linespacing\@plus.7\linespacing}{-.5em}%
  {\normalfont\bfseries}}
\def\paragraph{\@startsection{paragraph}{4}%
  \z@\z@{-\fontdimen2\font}%
  \normalfont\bfseries}
\def\subparagraph{\@startsection{subparagraph}{5}%
  \z@\z@{-\fontdimen2\font}%
  \normalfont\bfseries}

\makeatother

\newcommand{\AT}{e^{-2i\pi \frac{c}{24}}}

\newcommand{\confbl}{\mathscr{F}}

\newcommand{\Drinfeld}{\boldsymbol{\chi}}

\newcommand{\Radford}{\pmb{\phi}{}}
\newcommand{\vvarkappa}{\pmb{\boldsymbol{\varkappa}}}

\newcommand{\hrho}{\widehat{\boldsymbol{\rho}}}

\newcommand{\bomega}{\boldsymbol{\Omega}}

\newcommand{\radmap}{\widehat{\pmb{\boldsymbol{\phi}}}}

\newcommand{\ff}{\boldsymbol f}
\newcommand{\bg}{\boldsymbol g}

\newcommand{\cZ}{\mathsf{Z}}
\newcommand{\ZisoZ}{\mathcal{F}}

\newcommand{\andp}{\relax}

\newcommand{\UresSL}[1]{\overline{\mathscr{U}}_{\q} s\ell(#1)} 
\newcommand{\tqalgA}{\overline{\mathscr{U}}_{\q} s\ell(2)}

\newcommand{\voal}[1]{\boldsymbol{\mathscr{#1}}} 
\newcommand{\algW}{\voal{W}}

\newcommand{\rep}{\mathcal}  

\newcommand{\repX}{\rep{X}}

\newcommand{\repP}{\rep{P}}

\newcommand{\XX}{\mathsf{X}} 

\newcommand{\XXgen}{\mathsf{X}}

\newcommand{\PP}{\mathsf{P}}

\newcommand{\ZZ}{\mathsf{Z}}
\newcommand{\GG}{\mathsf{G}}
\newcommand{\ZZcft}{\mathsf{Z}_\mathsf{cft}}

\renewcommand{\geq}{\,{\geqslant}\,}
\renewcommand{\leq}{\,{\leqslant}\,}
\newcommand{\tensor}{\otimes}

\newcommand{\cket}[1]{|\kern-1pt|#1\rangle\!\rangle}
\newcommand{\cbra}[1]{\langle\!\langle{}#1|\kern-1pt|}
\newcommand{\iket}[1]{|#1\rangle\!\rangle}
\newcommand{\ibra}[1]{\langle\!\langle{}#1|}
\newcommand{\half}{%
  \mathchoice{\ffrac{1}{2}}{\frac{1}{2}}{\frac{1}{2}}{\frac{1}{2}}}
\newcommand{\rmi}{{\rm i}}
\newcommand{\dd}{\partial}
\newcommand{\one}{\boldsymbol{1}}
\newcommand{\Log}{\pmb{\boldsymbol{{1}\kern-4pt1}}}
\newcommand{\Tr}{\mathrm{Tr}^{\vphantom{y}}}
\newcommand{\tr}{\mathrm{Tr}^{\vphantom{y}}}

\newcommand{\qTr}{\mathrm{qCh}}
\newcommand{\ad}{\mathrm{Ad}}
\newcommand{\id}{\mathrm{id}}

\newcommand{\Ch}{\mathsf{Ch}}

\newcommand{\balance}{{\boldsymbol{g}}}
\newcommand{\coint}{{\boldsymbol{c}}}
\newcommand{\rint}{{\boldsymbol{\mu}}}
\newcommand{\ribbon}{{\boldsymbol{v}}}

\newcommand{\cas}{\boldsymbol{C}}

\newcommand{\idem}{\boldsymbol{e}}
\newcommand{\nilp}{\boldsymbol{w}}

\newcommand{\mfrac}[2]{\mbox{\small$\displaystyle\frac{#1}{#2}$}}

\newcommand{\toppr}{\mathsf{t}}
\newcommand{\botpr}{\mathsf{b}}
\newcommand{\leftpr}{\mathsf{l}}
\newcommand{\rightpr}{\mathsf{r}}

\newcommand{\stprp}{\mathsf{a}}

\newcommand{\Bnm}{\mathsf{B}}
\newcommand{\Tnm}{\mathsf{T}}

\newcommand{\SLiiZ}{SL(2,\oZ)}
\newcommand{\hSL}[1]{\widehat{s\ell}(#1)}

\newcommand{\modS}{\mathscr{S}}

\newcommand{\Svac}{S^{\text{vac}}}
\newcommand{\modT}{\mathscr{T}}

\newcommand{\cycla}{\mathsf{a}}
\newcommand{\cyclb}{\mathsf{b}}

\newcommand{\smatrix}[4]{\left(\begin{smallmatrix}#1&#2\\
      #3&#4\end{smallmatrix}\right)}

\newcommand{\oC}{\mathbb{C}}
\newcommand{\oN}{\mathbb{N}}
\newcommand{\oZ}{\mathbb{Z}}

\newcommand{\bref}[1]{\textbf{\ref{#1}}}
\newcommand{\ffrac}[2]{\mbox{\footnotesize$\displaystyle\frac{#1}{#2}$}}

\renewcommand{\tilde}{\widetilde}

\newcommand{\q}{\mathfrak{q}}

\newcommand{\qbin}[2]{\mathchoice%
  {{\qbinm{#1}{#2}}}{\qbinmm{#1}{#2}}%
  {\qbinmm{#1}{#2}}{\qbinmm{#1}{#2}}}
\newcommand{\qbinm}[2]{\mbox{\footnotesize$\displaystyle
    \genfrac{[}{]}{0pt}{}{#1}{#2}$}}
\newcommand{\qbinmm}[2]{\genfrac{[}{]}{0pt}{}{#1}{#2}}

\swapnumbers
\newtheorem{Thm}[subsection]{Theorem}
\newtheorem{thm}[subsubsection]{Theorem}

\newtheorem{lemma}[subsubsection]{Lemma}
\newtheorem{Prop}[subsection]{Proposition}
\newtheorem{prop}[subsubsection]{Proposition}

\theoremstyle{definition}

\newtheorem{rem}[subsubsection]{Remark}

\begin{document}

\title[Radford, Drinfeld, and Cardy boundary states in $(1,p)$ models]
{
  Radford, Drinfeld, and Cardy boundary states in $(1,p)$ logarithmic
  conformal field models}

\author[Gainutdinov]{A.M.~Gainutdinov}%

\address{\mbox{}\kern-\parindent  
  amg, iyt: Division of Theoretical Physics, Lebedev Physical Institute,
  Leninsky prospekt 53, Moscow 119991, Russia 
  \hfill\mbox{}\linebreak 
  \texttt{azot@mccme.ru, tipunin@td.lpi.ru}}

\author[Tipunin]{I.Yu.~Tipunin}

\begin{abstract}
  We introduce $p-1$ pseudocharacters in the space of $(1,p)$ model
  vacuum torus amplitudes to complete the distinguished basis in the
  $2p$-dimensional fusion algebra to a basis in the whole
  $(3p-1)$-dimensional space of torus amplitudes, and the structure
  constants in this basis are integer numbers.  We obtain a generalized
  Verlinde-formula that gives these structure constants.
  In the context of theories with boundaries, we identify the space of
  vacuum torus amplitudes with the space of Ishibashi states.  Then,
  we propose $3p-1$ boundary states satisfying the Cardy condition.
\end{abstract}

\maketitle

\thispagestyle{empty}

\setcounter{tocdepth}{2}

\vspace*{-24pt}

\begin{scriptsize}
  \renewcommand{\textbf}{\relax}
  \addtolength{\baselineskip}{-12pt}
  \tableofcontents
\end{scriptsize}

\section{Introduction}
{\bf 1.0.}  Logarithmic conformal field
theories~\cite{[GK2],[PRZ],[S-ABC],[GaberdielKausch3],[FGST4],[GL],[S-sl2],[FFHST],[EF]}
attracts attention because their significance for applications (like
sand-pile model~\cite{[Ruelle], [Jeng]} and percolation~\cite{[Cardy],
[Watts], [DuplantierSaleur]}) as well as for general questions in the
theory itself~\cite{[HLZ],[AM]}.  Between logarithmic conformal field
theories the $(1,p)$ models~\cite{[Flohr],[G-alg]} are studied in most
detail~\cite{[GaberdielKausch], [GaberdielKausch3], [GK2]}. The
$(1,p)$ models are characterized by the central charge
\begin{equation}\label{centr-charge}
  c=13-6(p + p^{-1}),
\end{equation}
and the spectrum of conformal dimensions of primary fields is given by
\begin{equation*}
  \Delta_r=\ffrac{r^2-1}{4p} + \ffrac{1-r}{2}, \qquad 1\leq r\leq p.
\end{equation*}
The chiral symmetry algebra of the $(1,p)$ models is the triplet
$W$-~algebra~\cite{[K-first]}, which we denote in what follows as~$\algW_p$. 

Logarithmic conformal field theories with boundaries are of great
importance in considering their applications in lattice models, which
usually involve boundaries~\cite{[JPR],[PRZ],[IPRH]}. Various attempts
to understand boundary logarithmic theories have been made in the
past~\cite{[Kaw],[KogW],[Ish],[BredF],[CQS], [Bred]} and recently
\cite{[GR1],[GR2]}.

In this paper we propose using the Kazhdan--Lusztig correspondence
stated for some logarithmic models~\cite{[FGST],[FGST3],[FGST4]} in
studying the boundary theories. The Kazhdan--Lusztig correspondence
for the $(1,p)$ models is an equivalence~\cite{[FGST2]} between
representation categories of the triplet algebra $\algW_p$ and the
\textit{restricted} quantum group $\UresSL2$, with
$\q=e^{\frac{\rmi\pi}{p}}$. Strictly speaking, these categories are
equivalent as ribbon braided quasitensor categories
(see~\cite{[FGST2],[S-q]}). Such equivalence leads to the following
identifications:
\begin{enumerate}
\item
the center $\ZZcft$ of the $\algW_p$ representation category and the
center $\ZZ$ of $\UresSL2$ are isomorphic as associative commutative algebras;
\item
the modular group actions on $\ZZcft$ and $\ZZ$ are equivalent.
\end{enumerate}
The space $\ZZcft$ is spanned by central elements (or endomorphisms)
of the bimodule structure 
on the full space of bulk states; we mean the bimodule structure with
respect to the chiral and antichiral actions. As in rational
cases~\cite{[BPPZ]}, in the context of the boundary theories, these
endomorphisms can be identified with boundary states. Therefore,
keeping in mind the identifications (1) and (2), the center $\ZZ$ of
$\UresSL2$ and some additional structures (multiplication, modular
group action) on it allow describing the space of boundary states for
the $(1,p)$ models.

\medskip
\addtocounter{subsection}{1}
\newcounter{indsec}
\setcounter{indsec}{\arabic{subsection}}

\noindent
\textbf{\thesubsection. \ Our main results.}
The space $\ZZcft$ is $(3p-1)$-dimensional
and has a distinguished basis of the vacuum torus amplitudes, i.e.,
of the characters
\begin{equation*}
  \chi_{s}^{\pm}(\tau)
  = \Tr_{\repX^{\pm}(s)}q^{L_0-\frac{c}{24}}, 
  \qquad 1\leq s\leq p,
\end{equation*}
 and the pseudocharacters
\begin{equation*}
  \chi_{s}(\tau) 
  = \Tr_{\repP^+(s)\oplus\repP^-(p-s)}(q^{L_0-\frac{c}{24}} \Log), \quad
  1\leq s\leq p-1,
\end{equation*}
where $\Log$ is the logarithmic partner of the identity operator
$\one$, and $\repP^{\pm}(s)$ are the projective covers of the
irreducible $\algW_p$-representations $\repX^{\pm}(s)$; 
here, we also set $q=e^{2\rmi\pi \tau}$, where $\tau$ is the
modular parameter.
The space $\ZZcft$ can be endowed with a commutative associative
algebra structure, which we introduce below
in~\bref{sec:mult-cft}. The structure constants in $\ZZcft$ in the
basis of the characters and pseudocharacters are \textit{integer}
numbers.
\begin{prop}\label{prop-intro:chi-mult}
For $1\leq r,s \leq p$, and $\alpha,\beta=\pm$,
\begin{equation}\label{chi-fusion}
  \chi^{\alpha}_r\chi^{\beta}_s
  =\sum_{\substack{t=|r - s| + 1\\
      \mathrm{step}=2}}^{r + s - 1}
  \tilde\chi^{\alpha\beta}_t,
\qquad  \tilde\chi^{\alpha}_t
  =
  \begin{cases}
    \chi^{\alpha}_t,&1\leq t\leq p,\\
    \chi^{\alpha}_{2p - t} + 2\chi^{-\alpha}_{t - p}, & p + 1 \leq t
    \leq 2p - 1,
  \end{cases}
\end{equation}
and, for $1\leq r,s \leq p-1$,
\begin{gather*}
 \chi^+_r\chi_s = \sum_{\substack{t=|r-s|+1\\
  \mathrm{step}=2}}^{\substack{\mathrm{min}(r+s-1,\\2p-r-s-1)}} \chi_t\;,\qquad
 \chi^-_{p-r}\chi_s = -\!\!\!\sum_{\substack{t=|r-s|+1\\
  \mathrm{step}=2}}^{\substack{\mathrm{min}(r+s-1,\\2p-r-s-1)}} \chi_t\;,
\qquad \chi^{\pm}_p\chi_s = 0,\qquad
\chi_r\chi_s=0,
\end{gather*}
where we set $\chi^{\pm}_s=\chi^{\pm}_s(\tau)$ and
$\chi_s=\chi_s(\tau)$ for simplicity.
\end{prop}
The multiplication~\eqref{chi-fusion} was first derived
in~\cite{[FHST]} as fusion of irreducible
$\algW_p$-re\-pre\-sen\-ta\-tions and was subsequently shown
in~\cite{[FGST]} to be the Grothendieck ring of~$\UresSL2$. The
elements $\chi^{\pm}_s$ span the fusion algebra $\GG\subset\ZZcft$;
the pseudocharacters $\chi_s$ span a $(p-1)$-dimensional ideal
in~$\ZZcft$.

The space $\ZZcft$ admits the modular group action generated by
\begin{equation*}
S:\tau\mapsto -1/\tau\;,\qquad 
T:\tau\mapsto\tau+1.
\end{equation*}  

The structure constants in~\bref{prop-intro:chi-mult}
are reproduced from the $S$-matrix action and we thus get a
generalized Verlinde-formula for the $(1,p)$ models.
\begin{prop}
The structure constants in $\ZZcft$ with respect to the basis of the
characters and pseudocharacters are given by
\begin{equation}\label{fusion-coef-intro}
N_{[r;\alpha][s;\beta]}^{[k;\gamma]}
=\sum_{l=1}^{p+1}\sum_{\lambda=1}^{n_l}
\ffrac{\Svac_{[l;1]} S^{[r;\alpha]}_{[l;1]}
S^{[s;\beta]}_{[l;\lambda]} +
\Svac_{[l;1]}S^{[r;\alpha]}_{[l;\lambda]}
S^{[s;\beta]}_{[l;1]}-\Svac_{[l;\lambda]}S^{[r;\alpha]}_{[l;1]}
S^{[s;\beta]}_{[l;1]}}{\bigl(\Svac_{[l;1]}\bigr)^2}S^{[l;\lambda]}_{[k;\gamma]},
\end{equation}
where $1\leq r,s,k\leq p+1$,
$1\leq\alpha,\beta,\gamma\leq n_l\leq3$, and $\Svac_{[r;\alpha]}$ are the
``vacuum'' row elements
(the notations are introduced below in Sec.~\ref{sec:mod-act-tor-ampl}
and Sec.~\ref{sec:gen-Verlinde}).
\end{prop}

The Verlinde formula \eqref{fusion-coef-intro} can be considered as a
generalization of $(1,p)$ model Verlinde formulas derived
in~\cite{[FHST],[GR2]} because \eqref{fusion-coef-intro} gives the
structure constants in the whole space of torus amplitudes, in which
the fusion algebra is a $2p$-dimensional subalgebra.

The generalized Verlinde-formula~\eqref{fusion-coef-intro} is based on
the following findings. The multiplication in $\ZZcft$ in the basis 
\begin{equation}\label{basis-phi}
\phi^{\pm}_s = S(\chi^{\pm}_s),\qquad
\phi_s = S(\chi_s),
\end{equation}
is block-diagonal, and the structure constants are expressed in terms
of the $S$ matrix vacuum row elements.
\begin{prop}\label{prop-intro:phi-mult}
 For $1\leq r,s\leq p-1$, the only nonzero multiplications in $\ZZcft$
with respect to the basis~\eqref{basis-phi} are given by
\begin{align*}
&\phi_r\,\phi_r = \; \ffrac{1}{\Svac_{[r;1]}} \bigl(
 \phi_r - \ffrac{\Svac_{[r;2]}}{\Svac_{[r;1]}}\phi^{+}_r 
-\ffrac{\Svac_{[r;3]}}{\Svac_{[r;1]}}\phi^{-}_{p-r}\bigr),\\
&\phi_r\,\phi^+_r = \; \ffrac{1}{\Svac_{[r;1]}}\, \phi^+_r,\quad
\phi_r\,\phi^-_{p-r} = \; \ffrac{1}{\Svac_{[r;1]}}\, \phi^-_{p-r},\\
&\phi^{+}_p\,\phi^{+}_p = \; \ffrac{1}{\Svac_{[p;1]}}\, \phi^{+}_p,\quad
\phi^{-}_p\,\phi^{-}_p = \; \ffrac{1}{\Svac_{[p+1;1]}}\,
\phi^{-}_p.
\end{align*}
Here, $\Svac_{[r;\alpha]}$, with $\alpha=1,2,3$, are the ``vacuum''
row elements (the notations are introduced below in
Sec.~\ref{sec:mod-act-tor-ampl}).
\end{prop}

\medskip
\addtocounter{subsubsection}{1}
\newcounter{indsec1}
\setcounter{indsec1}{\arabic{subsubsection}}

\noindent
\textbf{\thesubsubsection. \ Boundary states.}  We identify the space
of Ishibashi states with the space of vacuum torus amplitudes
$\ZZcft$.  Then, we propose $3p-1$ Cardy states that formally satisfy
the (extended) Cardy condition
 \begin{equation*}
  \cbra{[r;\alpha]}q^{\half(L_0+\bar
  L_0-\frac{c}{12})}\cket{[s;\beta]}=\sum_{k=1}^{p+1}
  \sum_{\gamma=1}^{n_k} N_{[r;\alpha][s;\beta]}^{[k;\gamma]}
  \chi_{[k;\gamma]}(\tilde q),
\end{equation*}
where $\tilde q = e^{-2\rmi\pi/\tau}$ and the integer structure
constants $N_{[r;\alpha][s;\beta]}^{[k;\gamma]}$ are given
in~\eqref{fusion-coef-intro}, with the notations introduced in
Sec.~\ref{sec:mod-act-tor-ampl} and Sec.~\ref{sec:boundary-st}.
 
 The paper is organized as follows. In
 Sec.~\ref{sec:mod-act-tor-ampl}, we introduce the space $\ZZcft$ of
 vacuum torus amplitudes for the $(1,p)$ models and recall the modular
 group action on~$\ZZcft$ in~\bref{sec:S-matrix-cft}, closely
 following to~\cite{[FHST],[FGST]}.  In~\bref{sec:mult-cft}, we
 introduce an associative algebra structure in the space of torus
 amplitudes. Sec.~\ref{sec:the-quant} is designed to compute the
 multiplications in $\ZZcft$ using some quantum-group techniques.
 In~\bref{subsec:def-qg}, we recall the quantum group $\UresSL2$ dual
 to the triplet algebra $\algW_p$. In~\bref{thm:center-mult}, we
 calculate the multiplications in the center~$\ZZ$ with respect to two
 distinguished bases related by the $S$-transformation from the
 modular group and we thus obtain the multiplications in $\ZZcft$. In
 Sec.~\ref{sec:gen-Verlinde}, these results are then applied to obtain
 a generalized Verlinde formula for the $(1,p)$ models.  With this
 information we then analyze the boundary states for the $(1,p)$
 models in Sec.~\ref{sec:boundary-st}. Conclusions are given in
 Sec.~\ref{sec:concl}.  The Appendices contain auxiliary or bulky
 facts and proofs.

\subsection*{Notations}
We use the standard abuse of notation for characters: we write
 $\chi(\tau)$ for  $\chi(e^{2\rmi\pi\tau})$, and set in what follows
\begin{equation*}
q=e^{2\rmi\pi \tau},\qquad \tilde q = e^{-2\rmi\pi/\tau}.
\end{equation*}
 We set in the paper
\begin{equation*}
  \q=e^{\frac{\rmi\pi}{p}},
\end{equation*}
for an integer $p\geq2$, and use the standard notation
\begin{equation*}
  [n] = \ffrac{\q^n-\q^{-n}}{\q-\q^{-1}},\quad
  n\in\oZ,\quad [n]! = [1][2]\dots[n],\quad n\in\oN,\quad[0]!=1
\end{equation*}
(without indicating the ``base''~$\q$ explicitly) and set
\begin{equation*}
  \qbin{m}{n}=
  \begin{cases}
    0,& n<0\quad\text{or}\quad m-n<0,\\
    \ffrac{[m]!}{[n]!\,[m-n]!}&\text{otherwise}.
  \end{cases}
\end{equation*}

For Hopf algebras in general and for $\UresSL2$ specifically, we write
$\Delta$, $\epsilon$, and~$S$ for the comultiplication, counit, and
antipode respectively.

We write $x'$, $x''$ (Sweedler's notation) in
$\Delta(x)=\sum_{(x)}x'\tensor x''.$

For a linear function $\beta$, we use the notation $\beta(?)$, where
$?$ indicates the position of its argument in more complicated
constructions.

\section{The spase $\ZZcft$ of $(1,p)$ model torus
  amplitudes} \label{sec:mod-act-tor-ampl}

We briefly recall the definition of $(1,p)$ logarithmic models and
their chiral symmetry algebra in~\bref{subsec:log-model}.  We
introduce the space of vacuum torus amplitudes associated with the
$(1,p)$ logarithmic models in~\bref{subsec:torus-ampl} and recall
their modular properties in~\bref{sec:S-matrix-cft}. In
\bref{sec:mult-cft}, we introduce multiplication in the space of torus
amplitudes.

\subsection{Logarithmic $(1,p)$ models}\label{subsec:log-model}
Here, we closely follow to~\cite{[FHST]}. Logarithmic $(1,p)$ models
are defined as kernels of certain screening operators, which commute
with the Virasoro algebra.  The actual symmetry of the theory is the
maximal local algebra in this kernel. In a $(1,p)$-model, which is the
kernel of the ``short'' screening operator, see~\cite{[FHST]}, this is
the triplet W-algebra $\algW_p$ studied in~\cite{[K-first],[GK2]}. The
chiral algebra $\algW_p$ has $2p$ irreducible representations
$\repX^{\pm}(s)$ and $2p$ projective covers $\repP^{\pm}(s)$
\cite{[FHST]} of the irreducibles ($1\leq s\leq p$). The quantum-group
counterparts of these modules are defined in App.~\bref{app:irr-proj}.

\subsection{The spase of $(1,p)$ model torus amplitudes}\label{subsec:torus-ampl}
The space $\ZZcft$ of vacuum torus amplitudes is $3p-1$ dimensional
\footnote{see~\cite{[FG]} for the case $p=2$.}
and is spanned by $2p$ characters of irreducible representations
$\repX^{\pm}(s)$ of $\algW_p$,
\begin{equation}\label{char-def}
  \chi_{s\andp}^{+}(\tau)
  =\Tr_{\repX^+(s)}e^{2\rmi\pi\tau(L_0-\frac{c}{24})}, 
  \qquad\chi_{s\andp}^{-}(\tau)
  =\Tr_{\repX^-(s)}e^{2\rmi\pi\tau(L_0-\frac{c}{24})},
  \qquad
  1\leq s\leq p,
\end{equation}
given in~\eqref{eq:characters} in terms of theta functions, and $p-1$
pseudocharacters assigned to the projective modules
$\repP^+(s)\oplus\repP^-(p-s)$,
\begin{multline}\label{pseudochar-def}
  \chi_{s\andp}(\tau) =  a_0 \tau\bigl(\ffrac{p-s}{p}\,\chi_{s\andp}^{+}(\tau) -
  \ffrac{s}{p}\,\chi_{p-s\andp}^{-}(\tau)\bigr)=\\
  =\Tr_{\repP^+(s)\oplus\repP^-(p-s)}(e^{2\rmi\pi\tau(L_0-\frac{c}{24})}\Log), \quad
  1\leq s\leq p-1,
\end{multline}
where $\Log$ is the logarithmic partner of the vacuum field (identity
operator $\one$). The pseudocharacters $\chi_{s}(\tau)$ are given
in~\eqref{eq:pseudochar}.
\begin{rem}
We note that the pseudocharacters \eqref{pseudochar-def} have no
canonical normalization and are therefore defined up to a common nonzero
constant $a_0$. However, the constant $a_0$ does not appear in final results.
\end{rem}

We define a vector of the characters and the pseudocharacters
\begin{equation}\label{basis-chi-def-simple}
  (\underbrace{\chi_1(\tau),\chi^+_1(\tau),\chi^-_{p-1}(\tau),\dots,
  \chi_{p-1}(\tau),\chi^+_{p-1}(\tau),\chi^-_{1}(\tau)}_{3\times(p-1)},
  \chi^+_{p}(\tau),\chi^-_{p}(\tau)),
\end{equation}
where we group the characters and pseudocharacters into $p-1$
triplets $\{\chi_s(\tau),\chi^+_s(\tau),\chi^-_{p-s}(\tau)\}$, for $1\leq
s\leq p-1$, and into two singlets $\chi^+_{p}(\tau)$, and
$\chi^-_{p}(\tau)$.

In what follows, we use $2$-index parametrization of the elements
in~\eqref{basis-chi-def-simple},
\begin{equation}\label{basis-chi-def}
\begin{split}
\bigl(\chi_{[s;1]}(\tau), \chi_{[s;2]}(\tau), \chi_{[s;3]}(\tau)\bigr) 
&= \bigl(\chi_s(\tau),\chi^+_s(\tau),\chi^-_{p-s}(\tau)\bigr), \qquad
1\leq s\leq p-1,\\
\chi_{[p;1]}(\tau) = \chi^+_{p}(\tau), &\quad \chi_{[p+1;1]}(\tau) = \chi^-_{p}(\tau).
\end{split}
\end{equation}

\subsection{The modular group action on $\ZZcft$}\label{sec:S-matrix-cft}
The $\SLiiZ$ action in terms of the vector
$(\chi_{[s;\alpha]}(\tau))$ in~\eqref{basis-chi-def} can be written as follows
  \begin{equation*}
    \chi_{[s;\alpha]}(-1/\tau)=\sum_{j=1}^{p+1}\sum_{\beta=1}^{n_j}
    S_{[s;\alpha][j;\beta]} \chi_{[j;\beta]}(\tau),\qquad 1\leq s\leq p-1,
  \end{equation*}
where $n_j=3$, for $1\leq j\leq p-1$, and $n_j=1$, for $j=p,p+1$, and
the matrix $S_{[s;\alpha][j;\beta]}$ has the following block
structure~\cite{[FGST]} (see also \cite{[FHST],[F-95]})
  \begin{gather}\label{S-block}
    \begin{matrix}
      3\times3 &\dots & 3\times3 & 3\times2\\
      \vdots &\ddots &\vdots  & \vdots\\
      3\times3 &\dots & 3\times3 & 3\times2 \\
      2\times3 &\dots &2\times3 & 2\times2 \\
    \end{matrix}\;,
  \end{gather}
  where the $3\times3$ blocks, labeled by $(s,j)$ with
  $s,j=1,\dots,p-1$, are given by
  \begin{multline}\label{S-3x3}
    \begin{pmatrix}
       S_{[s;1][j;1]} & S_{[s;1][j;2]}  & S_{[s;1][j;3]}\\
       S_{[s;2][j;1]} & S_{[s;2][j;2]}  & S_{[s;2][j;3]}\\
       S_{[s;3][j;1]} & S_{[s;3][j;2]}  & S_{[s;3][j;3]}
    \end{pmatrix}=\\
    =  \ffrac{(-1)^{p+s+j}}{\sqrt{2p}}
    \begin{pmatrix}
      0
      &a_0\ffrac{p-j}{p}(\q^{sj} - \q^{-sj}) 
      & -a_0\ffrac{j}{p}(\q^{sj} - \q^{-sj})\\[6pt]
      -\ffrac{1}{a_0}(\q^{sj} - \q^{-sj})
      &\ffrac{s}{p}(\q^{sj} + \q^{-sj}) 
      & \ffrac{s}{p}(\q^{sj} + \q^{-sj})\\[6pt]
      \ffrac{1}{a_0}(\q^{sj} - \q^{-sj})
      & \ffrac{p-s}{p}(\q^{sj} + \q^{-sj}) 
      & \ffrac{p-s}{p}(\q^{sj} + \q^{-sj})
    \end{pmatrix},
  \end{multline}
  with $\q=e^{\frac{\rmi\pi}{p}}$, the $3\times2$ blocks, labeled by
  $(s,p)$ with $s=1,\dots,p-1$, are
  \begin{equation}\label{S-3x2}
    \begin{pmatrix}
       S_{[s;1][p;1]} & S_{[s;1][p+1;1]}\\
       S_{[s;2][p;1]} & S_{[s;2][p+1;1]}\\
       S_{[s;3][p;1]} & S_{[s;3][p+1;1]}
    \end{pmatrix}=
    \ffrac{1}{p\sqrt{2p}}\begin{pmatrix}
      0\;&0\\
      s\;&(-1)^{p-s}s\\
      p-s\;&(-1)^{p-s}(p-s)
    \end{pmatrix},
  \end{equation}
  the $2\times3$ blocks, labeled by $(p,j)$ with $j=1,\dots,p-1$,
  are
  \begin{equation}\label{S-2x3}
    \begin{pmatrix}
       S_{[p;1][j;1]} & S_{[p;1][j;2]} & S_{[p;1][j;3]}\\
       S_{[p+1;1][j;1]} & S_{[p+1;1][j;2]} & S_{[p+1;1][j;3]}
    \end{pmatrix}=
    \ffrac{2}{\sqrt{2p}}\begin{pmatrix}
      0\;&1\;&1\\
      0\;&(-1)^{p-j}\;&(-1)^{p-j}
    \end{pmatrix},
  \end{equation}
  and the $2\times2$ block is given by
  \begin{equation}\label{S-2x2}
    \begin{pmatrix}
       S_{[p;1][p;1]} & S_{[p;1][p+1;1]}\\
       S_{[p+1;1][p;1]} & S_{[p+1;1][p+1;1]}
    \end{pmatrix}=
    \ffrac{1}{\sqrt{2p}}\begin{pmatrix}
      1\;&1\\
      1\;&(-1)^p
    \end{pmatrix}.
  \end{equation}

The $S$ matrix has a distinguished row 
\begin{equation}\label{Svac-def}
(\Svac_{[j;\beta]})=(S_{[1;2][j;\beta]}),
\end{equation}
which corresponds to the $S$-transformation of the vacuum
representation character $\chi^+_1(\tau)$:
\begin{equation*}
\chi^+_1(-1/\tau) = \sum_{j=1}^{p+1}\sum_{\beta=1}^{n_j}
\Svac_{[j;\beta]} \chi_{[j;\beta]}(\tau).
\end{equation*}

The $T$-transformation on the space of torus amplitudes is given
in~\cite{[FHST]} and we do not reproduce the $T$-action here. In what
follows, we need the properties of $\ZZcft$ with respect to the
$S$-transformation only.

\subsection{Multiplication in the space of torus amplitudes}\label{sec:mult-cft}
The space of vacuum torus amplitudes can be endowed with an
associative commutative algebra structure in the way similar to the one
in~\cite{[MS1],[MS2]} for semisimple (rational) cases. Here, we
introduce such algebra structure on the space $\ZZcft$ of torus
amplitudes for the $(1,p)$ models. But the reader should note that we
give only heuristic description.

Let $\cycla$ and $\cyclb$ denote two basic Dehn twists in a torus
depicted on the Fig.~\eqref{torus-pic}.
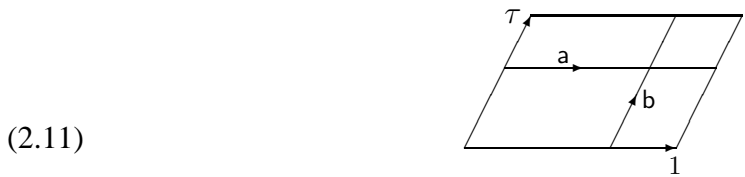
\begin{equation}\label{torus-pic}
\begin{picture}(80,60)
\put(0,0){\vector(1,2){25}}
\put(15,47){$\tau$}             
\put(0,0){\vector(1,0){80}}
\put(77,-10){\footnotesize$1$}   
\put(25,50){\line(1,0){80}}
\put(80,0){\line(1,2){25}}
\put(15,30){\vector(1,0){30}}
\put(45,30){\line(1,0){50}}
\put(35,32){\footnotesize$\cycla$}  
\put(55,0){\vector(1,2){10}}
\put(65,20){\line(1,2){15}}
\put(67,15){\footnotesize$\cyclb$}  
\end{picture}
\end{equation}
where $\tau$ is in the upper-half plane.  The characters
$\chi^{\pm}_s(\tau)$ in~\eqref{char-def} and the pseudocharacters
$\chi_s(\tau)$ in~\eqref{pseudochar-def} are conformal blocks of
zero-point correlations on the torus. We depict these conformal blocks
as
\begin{equation*}
  \begin{picture}(110,50)
    \put(-35,27){$\chi^{\pm}_s(\tau) = $}
    \put(30,30){\circle{30}}
    \put(33,14){\vector(1,0){0}}
    \put(40,10){\footnotesize$\chi^{\pm}_s$}
  \end{picture}\qquad
  \begin{picture}(110,50)
    \put(-35,27){$\chi_s(\tau) = $}
    \put(30,30){\circle{30}}
    \put(33,14){\vector(1,0){0}}
    \put(15,30){\oval(9,13)}
    \put(11,26){$\Log$}
    \put(40,10){\footnotesize$\chi_s$}
  \end{picture}
\end{equation*}
with the cycles corresponded to the $\cyclb$ cycle in the torus. Next,
we introduce the conformal blocks 
\begin{equation*}
  \qquad\qquad
  \begin{picture}(110,100)
    \put(-68,27){$\confbl_{r,r*}^{\pm,s}(z-w) = $}
    \put(30,30){\circle{30}}
    \put(33,14){\vector(1,0){0}}
    \put(40,10){\footnotesize$\chi^{\pm}_s$}
    \put(30,46){\line(0,1){15}}
    \put(32,53){\small$\one$}
    \put(30,61){\line(1,1){14}}
    \put(30,61){\line(-1,1){14}}
    \put(2,80){\footnotesize$\psi_r(z)$}
    \put(45,80){\footnotesize$\psi_{r^*}(w)$}
  \end{picture}
\qquad\qquad\quad
  \begin{picture}(110,100)
    \put(-70,27){$\confbl_{r,r*}^{s}(z-w) = $}
    \put(30,30){\circle{30}}
    \put(33,14){\vector(1,0){0}}
    \put(15,30){\oval(9,13)}
    \put(11,26){$\Log$}
    \put(40,10){\footnotesize$\chi_s$}
    \put(30,46){\line(0,1){15}}
    \put(32,53){\small$\one$}
    \put(30,61){\line(1,1){14}}
    \put(2,80){\footnotesize$\psi_r(z)$}
    \put(30,61){\line(-1,1){14}}
    \put(45,80){\footnotesize$\psi_{r^*}(w)$}
  \end{picture}
\end{equation*}
for two-point correlators
$\langle\psi_r(z)\psi_{r^*}(w)\rangle$ with selfconjugate primary
fields $\psi_r(z)$ ($r=r^*$) with respect to $\algW_p$. The characters
$\chi^{\pm}_s(\tau)$ and the pseudocharacters $\chi_s(\tau)$ can be
obtained from these conformal blocks in the limit of coinciding
points,
\begin{equation*}
\chi^{\pm}_s(\tau) = \lim_{z\to
  w}(z-w)^{\Delta_r}\confbl_{r,r*}^{\pm,s}(z-w),\quad 
\chi_s(\tau) = \lim_{z\to
  w}(z-w)^{\Delta_r}\confbl_{r,r*}^{s}(z-w),
\end{equation*}
where $\Delta_r$ is the conformal dimension of $\psi_r(z)$.

We define multiplication ``$\star$'' of a primary field $\psi_r(z)$ with
$\chi^{\pm}_s(\tau)$ and $\chi_s(\tau)$ as the result of computing the
monodromy, along the $\cyclb$ cycle, of the conformal blocks
$\confbl_{r,r*}^{\pm,s}(z-w)$, and $\confbl_{r,r*}^{s}(z-w)$ with
respect to $z$ coordinate:
\begin{equation*}
  \qquad\qquad\qquad\qquad\quad
  \begin{picture}(110,100)
    \put(-100,30){$\psi_r(z)\star\chi^{\pm}_s(\tau) = $}
    \put(30,30){\circle{60}}
    \put(33,10){\vector(1,0){0}}
    \put(50,10){$\chi^{\pm}_s$}
    \put(30,50){\line(0,1){15}}
    \put(32,53){\small$\one$}
    \put(30,65){\line(1,1){18}}
    \put(30,65){\line(-1,1){18}}
    \put(2,90){\footnotesize$\psi_r(z)$}
    \put(45,90){\footnotesize$\psi_{r^*}(w)$}
    {\linethickness{0.2pt}
    \qbezier(7,83)(-40,40)(10,5)
    \qbezier(10,5)(30,-10)(60,0)
    \qbezier(60,0)(110,25)(43,75)
    \qbezier(40,77)(30,83)(20,83)}
    \put(20,83){\vector(-1,0){0}}
    \put(65,60){\small$\cyclb$}
  \end{picture}
\qquad\qquad\qquad\qquad\quad
  \begin{picture}(110,100)
    \put(-95,30){$\psi_r(z)\star\chi_s(\tau) = $}
    \put(30,30){\circle{60}}
    \put(10,30){\oval(9,13)}
    \put(6,26){$\Log$}
    \put(33,10){\vector(1,0){0}}
    \put(50,10){$\chi_s$}
    \put(30,50){\line(0,1){15}}
    \put(32,53){\small$\one$}
    \put(30,65){\line(1,1){18}}
    \put(30,65){\line(-1,1){18}}
    \put(2,90){\footnotesize$\psi_r(z)$}
    \put(45,90){\footnotesize$\psi_{r^*}(w)$}
    {\linethickness{0.2pt}
    \qbezier(7,83)(-40,40)(10,5)
    \qbezier(10,5)(30,-10)(60,0)
    \qbezier(60,0)(110,25)(43,75)
    \qbezier(40,77)(30,83)(20,83)}
    \put(20,83){\vector(-1,0){0}}
    \put(65,60){\small$\cyclb$}
  \end{picture}
\end{equation*}
and taking the limit of coinciding points at the end of the
computation.  Then, the multiplication in the space $\ZZcft$ of vacuum
torus amplitudes is defined as follows
\begin{equation*}
\chi^{\pm}_r\chi^{\alpha}_s = \psi^{\pm}_r(z)\star \chi^{\alpha}_s(\tau), \quad 
\chi^{\pm}_r\chi_s = \psi^{\pm}_r(z) \star\chi_s(\tau), \quad
\chi_r\chi_s = \psi_r(z)\star\chi_s(\tau),
\end{equation*}
where $\alpha\in\{+,-\}$, $\psi^{\pm}_r(z)$ are the primary fields of
$\repX^{\pm}(r)$, and $\psi_r(z)$ are appropriate logarithmic partners.

The Kazhdan--Lusztig correspondence stated in~\cite{[FGST]} for the
$(1,p)$ logarithmic models have lead in~\cite{[FGST2]} to an
equivalence between representation categories of the chiral algebra
$\algW_p$ and of $\UresSL2$. Such remarkable correspondence suggests
an isomorphism between the space $\ZZcft$ of torus amplitudes in the
$(1,p)$ models and the center $\ZZ$ of $\UresSL2$ as commutative
associative algebras. We compute the introduced multiplication in
$\ZZcft$ below in~\bref{thm:center-mult} using a quantum-group
approach.

\section{The quantum group center $\ZZ$}\label{sec:the-quant}
To describe $\ZZcft$ as an associative commutative algebra, we first
recall some facts about the center $\ZZ$ of the restricted quantum
group $\UresSL2$.  As was shown in~\cite{[FGST]}, the center $\ZZ$ is
$(3p-1)$-dimensional and admits an $\SLiiZ$-action in the following
way. The space $\ZZ$ contains two special bases. The first one
consists of the images of characters and pseudocharacters (those
associated with some indecomposable representations) under the
Drinfeld mapping $\Drinfeld$~\cite{[Drinfeld]} and the second one
under the Radford mapping $\Radford$~\cite{[Rad]}.  We call these
bases \textit{the Drinfeld and Radford bases}, respectively. Then, the
$\modS$-transformation from the modular group maps each vector from
the Drinfeld basis to a vector from the Radford basis and vice
versa. Schematically, the modular group action on the center is given
by the diagrams
\begin{equation*}\modS:
   \xymatrix@=16pt{
     &\Ch  \ar[dl]_{\Drinfeld} \ar[dr]^{\Radford}&\\
     \cZ\ar@{<->}@/_/[rr]_{\modS}&&\cZ
}\qquad
\modT:\quad
   \xymatrix@=32pt{
     \cZ\ar@{->}[r]^{\ribbon}&\cZ\;,
}
\end{equation*}
where $\modS=\smatrix{0}{1}{-1}{0}$, $\modT=\smatrix{1}{1}{0}{1}$, and $\Ch$
is the space of characters and pseudocharacters, and $\ribbon$ is a
ribbon element.  

As we show below in~\bref{thm:center-mult}, the structure constants in
$\ZZ$ with respect to the Drinfeld basis are integer numbers while in
the Radford basis the multiplication is block-diagonal, and the
structure constants are expressed in terms of the $S$ matrix vacuum
row elements. We then use this information in
Sec.~\ref{sec:gen-Verlinde} to obtain the generalized
Verlinde-formula~\eqref{fusion-coef-intro} for the $(1,p)$ models.

\subsection{The definition of $\UresSL2$}\label{subsec:def-qg}
The quantum group dual to the $(1,p)$ logarithmic model with
the chiral algebra $\algW_p$ is the ``restricted'' quantum $s\ell(2)$
denoted as $\UresSL2$~\cite{[FGST]} with
\begin{equation*}
\q = e^{\frac{\rmi\pi}{p}}.
\end{equation*}
The three generators $E$, $F$, and $K$ satisfy the standard relations
for quantum $s\ell(2)$,
\begin{gather*}
  KEK^{-1}=\q^2E,\quad
  KFK^{-1}=\q^{-2}F,\\
  [E,F]=\ffrac{K-K^{-1}}{\q-\q^{-1}},
\end{gather*}
with some additional constraints,
\begin{equation*}
  E^{p}=F^{p}=0,\quad K^{2p}=\one,
\end{equation*}
and the Hopf-algebra structure is given by
\begin{gather*}
  \Delta(E)=\one\otimes E+E\otimes K,\quad
  \Delta(F)=K^{-1}\otimes F+F\otimes\one,\quad
  \Delta(K)=K\otimes K,\\
  \epsilon(E)=\epsilon(F)=0,\quad\epsilon(K)=1,\\
  S(E)=-EK^{-1},\quad  S(F)=-KF,\quad S(K)=K^{-1}.
\end{gather*}

The elements of the PBW-basis of $\UresSL2$ are enumerated as
$E^i\,K^j\,F^\ell$ with $0\!\leq\!i\leq\!p\,-~1$, $0\leq j\leq 2p-1$,
$0\leq\ell\leq p-1$, and its dimension is therefore~$2p^3$.

\subsection{The quantum group center}\label{subsec:center-qg}
Here, we describe the center $\cZ$ of $\UresSL2$ in terms of the
Drinfeld basis and the Radford basis. To construct these bases, we
first use irreducible and projective modules over $\UresSL2$ to
produce the space $\Ch=\Ch(\UresSL2)$ of $q$-characters, which is dual
to the center. The irreducible and projective modules are defined in
Appendix~\bref{app:irr-proj}.  Then, we obtain two distinguished bases
in the center related by the $\modS$-transformation. 
In order to be self-contained we begin by reviewing some standard
facts about the space $\Ch$ of $q$-characters, and the Radford and
Drinfeld maps, following~\cite{[FGST],[FGST3]}.

\subsubsection{The space of $q$-characters for $\UresSL2$}\label{sec:q-chars}
 For a Hopf algebra $A$, the space $\Ch=\Ch(A)$
of $q$-characters is defined as
\begin{multline}\label{Ch-def}
  \Ch(A)=\bigl\{\beta\in A^* \bigm| \ad^*_x(\beta)
  =\epsilon(x)\beta\quad \forall x\in A\bigr\}\\
  = \bigl\{\beta\in A^* \bigm| \beta(xy)=\beta\bigl(S^2(y)x\bigr)
  \quad \forall x,y\in A\bigr\},
\end{multline}
where the coadjoint action $\ad^*_a:A^*\to A^*$ is
$\ad^*_a(\beta)=\beta\bigl(\sum_{(a)} S(a')?a''\bigr)$, $a\in A$,
$\beta\in A^*$, and $S$ is the antipod.

In what follows, we need the so-called \textit{balancing element}
$\balance\in A$ that satisfies~\cite{[Drinfeld]}
\begin{equation}\label{bal-def}
  \Delta(\balance)=\balance\tensor\balance,\quad
  S^2(x)=\balance x\balance^{-1}
\end{equation}
for all $x\in A$.  For $A=\tqalgA$, $\balance=K^{p+1}$ (see~\cite{[FGST]}).

\subsubsection{Irreducible representation traces}
\label{subsec:func-gamma}
The space of $q$-characters contains a homomorphic image of the
Grothendieck ring under the $q$-trace: for any
$A$-module~$\XXgen$,
\begin{equation}\label{qCh}
  \qTr_{\XXgen} \equiv \tr_{\XXgen}(\balance^{-1}?)\in\Ch(A),
\end{equation}
where $\balance$ is the balancing element~\eqref{bal-def}.  For
$A=\tqalgA$, we thus have a $2p$-dimen\-sion\-al subspace in $\Ch$
spanned by $q$-traces over irreducible modules, i.e., by
\begin{align}\label{gamma-pm-def}
  &\gamma^{\pm}(s):x\mapsto\tr_{\XX^{\pm}_{s}}(\balance^{-1}x),
  \quad
  1\leq s\leq p,\\
  &\gamma^{\pm}(s)\in\Ch,\notag
\end{align}
with $\balance^{-1} = K^{p-1}$.

\subsubsection{Pseudotraces}\label{sec:pseudo}
The space of $q$-characters $\Ch(\tqalgA)$ is not exhausted by
$q$-traces over irreducible modules; it also contains ``pseudotraces''
associated with projective modules. To construct the pseudotraces
(they can be considered quantum group counterparts of pseudotraces
in~\cite{[My]} and of pseudocharacters in~\eqref{pseudochar-def}), we
closely follow the strategy proposed in~\cite{[FGST3]}. We first
consider the maps
\begin{equation}\label{sigma-def1}
\sigma_s:\PP^{+}_{s}\oplus\PP^{-}_{p-s}\to\PP^{+}_{s}\oplus\PP^{-}_{p-s},
\end{equation}
defined by its action on the corresponding basis vectors (see
\bref{module-L} and \bref{module-P}): $\sigma_s$ acts by zero on all
basis elements except
\begin{equation}\label{sigma-def2}
  \sigma_s:\botpr^{(\pm,s)}_n
  \mapsto\alpha^{\pm}_s\toppr^{(\pm,s)}_n
  + \beta^{\pm}_s\botpr^{(\pm,s)}_n,
\end{equation}
with the coefficients $\alpha^{\pm}_s$ and $\beta^{\pm}_s$ chosen
arbitrarily, that is, the map $\sigma_s$ has the diagonal part:
$\botpr^{(\pm,s)}_n \mapsto \beta^{\pm}_s\botpr^{(\pm,s)}_n$, and the
nondiagonal part: $\botpr^{(\pm,s)}_n \mapsto
\alpha^{\pm}_s\toppr^{(\pm,s)}_n$ corresponding to the action from the
bottom (the socle) to the top of the projective module
$\PP^{+}_{s}\oplus\PP^{-}_{p-s}$.

For any such $\sigma_s$, we now define a functional on $\tqalgA$ as
\begin{equation}\label{gamma-def}
  \gamma(s):
  x\mapsto \Tr_{\PP^{+}_{s}\oplus\PP^{-}_{p-s}}
  (\balance^{-1} x \sigma_s).
\end{equation}

\begin{prop}\label{prop:gammaUpUp}
  For $1\leq s \leq p-1$,
  \begin{equation*}\label{gamma-general}
    \gamma(s)\in\Ch
  \end{equation*}
  if and only if
  \begin{equation*}
    \alpha^{+}_s=\alpha^{-}_s
  \end{equation*}
\end{prop}
The proof is similar to the one in~\cite{[FGST3]}, Prop.~2.3.4.

\subsubsection{The $\gamma$ basis} The space $\Ch$ of
$q$-characters of $\UresSL2$ is spanned by the elements
\begin{gather}\label{gamma-basis-def}
\gamma(s),\quad
\gamma^{+}(s), \quad \gamma^{-}(p-s), \;\; 1\leq s\leq p-1,\qquad
\gamma^{+}(p),\quad \gamma^{-}(p),
\end{gather}
where $\gamma^{\pm}(s)$ are defined in~\eqref{gamma-pm-def}, and
$\gamma(s)$ are defined in~\eqref{gamma-def} with $\sigma_s$ in the
definition~\eqref{sigma-def2} fixed as
\begin{equation}\label{sigma-def3}
  \sigma_s:\botpr^{(\pm,s)}_n \mapsto\alpha_s\toppr^{(\pm,s)}_n,\qquad
  \alpha_s = -a_0
  \ffrac{[s]}{p(\q-\q^{-1})},
\end{equation}
where we fix the diagonal part of $\sigma_s$ as zero, and $a_0$ is the
normalization constant from~\eqref{pseudochar-def}. We thus have
$3p-1$ linear independent $q$-characters. In what follows, we call
these elements the \textit{$\gamma$ basis}.

\subsubsection{Radford map of $\Ch(\tqalgA)$}\label{subsec:rad-map-def}
For a Hopf algebra $A$ with the right integral $\rint$ and the
left--right cointegral~$\coint$ (see the definitions
in~\cite{[FGST]} and references therein), \textit{the Radford map} $\Radford:A^*\to
A$ and its inverse $\Radford^{-1}:A\to A^*$ are given by
\begin{equation}\label{radford-def}
  \Radford(\beta)
  =\sum_{(\coint)}\beta(\coint')\coint'',
  \quad
  \Radford^{-1}(x)=\rint(S(x)?).
\end{equation}

We now calculate the Radford map $\Radford:\Ch\to\cZ$ on the $\gamma$
basis~\eqref{gamma-basis-def} in $\Ch(\tqalgA)$ to obtain the
Radford-basis in the center of $\tqalgA$:
\begin{itemize}
\item 
The Radford map $\Radford$ on the irreducible representation traces
$\gamma^{\pm}(s)$ is given by
\begin{equation*}
\Radford(\gamma^{\pm}(s))=\Radford^{\pm}(s)=
\sum_{(\coint)}\Tr_{\XX^{\pm}_s} (K^{p-1}\coint')\coint'',
\end{equation*}
where the left--right cointegral $\coint$ is given by~\cite{[FGST]}
\begin{gather}\label{coint}
  \coint=\zeta\,F^{p-1}E^{p-1}\sum_{j=0}^{2p-1}K^j
\end{gather}
with the normalization
$\zeta=\sqrt{\frac{p}{2}}\,\frac{1}{([p-1]!)^2}$.
In~\cite{[FGST]}, $\Radford^{\pm}(s)$ were calculated in the PBW-basis,
  \begin{equation}\label{phi-explicit}
    \Radford^{\pm}( s) =
    \zeta\!\sum_{n=0}^{s - 1}
    \sum_{i=0}^{n}
    \sum_{j=0}^{2p - 1}\!(\pm 1)^{ i + j} ([i]!)^2
    \q^{j(s - 1 - 2n)}\!\qbin{s - n + i - 1}{i}\!\qbin{n}{i}
    F^{p - 1 - i} E^{p - 1 - i} K^j,
  \end{equation}
\item
The Radford map $\Radford$ on the pseudotraces $\gamma(s)$ is given by
\begin{equation*}
\Radford(\gamma(s))=\Radford(s)= \sum_{(\coint)}\Tr_{\PP^{+}_{s}
  \oplus \PP^{-}_{p-s}}(K^{p-1}\coint'\sigma_{s}) \coint'',
\end{equation*}
where the map $\sigma_s$ is defined in~\eqref{sigma-def1} and~\eqref{sigma-def3}.
In~\bref{app:Rad-gamma}, we evaluate $\Radford(s)$ as
\begin{multline}\label{Rad-PBW}
\Radford(s) =
 \alpha_s\zeta\sum_{m=0}^{p-2} \sum_{j=0}^{2p-1}\Bigl(\sum_{n=0}^{s-1}
 q^{j(s-1-2n)} \Bnm^+_{n,p-1-m}(s)\;+\\
 + \sum_{k=0}^{p-s-1}
 q^{j(-s-1-2k)} \Bnm^-_{k,p-1-m}(p-s)\Bigr)F^m E^m K^{j}, \qquad 1\leq s\leq p-1,
\end{multline}
with $\alpha_s$ given in~\eqref{sigma-def3}, and the
coefficients $\Bnm^+_{n,m}(s)$ and $\Bnm^-_{k,m}(p-s)$ are given
in~\eqref{Bnm-pm}, \eqref{Bnm-plus-2}, and \eqref{Bnm-min-2}.
\end{itemize}

\subsubsection{Drinfeld map of $\Ch(\tqalgA)$}
\label{subsec:drinf-map-def}
For a Hopf algebra $A$ with the nondegenerate $M$-matrix
($M=R_{21}R_{12}\in A\tensor A$), which satisfies the relations
\begin{equation*}
(\Delta\tensor\id)(M)=R_{32}M_{13}R_{23},\quad
 M\Delta(x)=\Delta(x) M \quad \forall x\in A,
\end{equation*}\pagebreak[3] 
\textit{the Drinfeld map}
$\Drinfeld:A^*\to A$ is defined by
\begin{equation}\label{drinfeld-def}
  \Drinfeld(\beta)=(\beta\tensor\id)M.
\end{equation}
In a Hopf algebra $A$ with the nondegenerate $M$-matrix, the
restriction of the Drinfeld map to the space $\Ch$ of $q$-characters
gives an isomorphism $\Ch(A)\xrightarrow{\simeq}\cZ(A)$ of associative
algebras~\cite{[Drinfeld]}.

We now calculate the Drinfeld map $\Drinfeld:\Ch\to\cZ$ on the
$\gamma$ basis~\eqref{gamma-basis-def} in $\Ch(\tqalgA)$ to obtain the
Drinfeld-basis in the center of $\tqalgA$:
\begin{itemize}
\item
The Drinfeld map $\Drinfeld$ on the irreducible representation traces
$\gamma^{\pm}(s)$ is given by
\begin{equation*}
\Drinfeld(\gamma^{\pm}(s))=\Drinfeld^{\pm}(s)= (\Tr_{\XX^{\pm}_{s}}\tensor \id)
    \bigl((K^{p-1}\tensor\one)\, M \bigr),
    \quad
    1\leq s\leq p,
\end{equation*}
where the $M$-matrix is given by~\cite{[FGST]}
\begin{multline}\label{M-def}
  M =\ffrac{1}{2p}
  \sum_{m=0}^{p-1}\sum_{n=0}^{p-1}
  \sum_{i=0}^{2p-1}\sum_{j=0}^{2p-1}
  \ffrac{(\q - \q^{-1})^{m + n}}{[m]! [n]!}\,
  \q^{m(m - 1)/2 + n(n - 1)/2}\\*
  \times \q^{- m^2 - m j + 2n j - 2n i - i j + m i} 
  F^{m} E^{n} K^{j}\tensor E^{m} F^{n} K^{i}.
\end{multline}
From~(\cite{[FGST]}, Prop.4.3.1), we have
\begin{multline}\label{the-cchi}
    \Drinfeld^{\alpha}(s) = \alpha^{p+1}(-1)^{s+1}\sum_{n=0}^{s-1}
    \sum_{m=0}^{n}
    (\q-\q^{-1})^{2m} \q^{-(m+1)(m+s-1-2n)}\times{}\\*
    {}\times\qbin{s-n+m-1}{m}
    \qbin{n}{m} E^m F^m K^{s-1+\beta p - 2n + m},
  \end{multline}
  where $1\leq s\leq p$, $\alpha=\pm1$, and we set $\beta=0$ if
  $\alpha=+1$ and $\beta=1$ if $\alpha=-1$.

\item
The Drinfeld map $\Drinfeld$ on the pseudotraces $\gamma(s)$ is given by
\begin{equation*}
\Drinfeld(\gamma(s))=\Drinfeld(s)= (\Tr_{\PP^{+}_{s}
  \oplus \PP^{-}_{p-s}}\!\!\!\tensor \id)
    \bigl((K^{p-1}\tensor\one)\, M (\sigma_{s}\tensor \id)\bigr),
    \quad
    1\leq s\leq p-1,
\end{equation*}
where the map $\sigma_s$ is defined in~\eqref{sigma-def1}
and~\eqref{sigma-def3}. From~\eqref{M-def}, we obtain $\Drinfeld(s)$
by direct calculation,
\begin{multline}\label{the-pscchi}
  \Drinfeld(s)=\alpha_s\sum_{m=1}^{p-1} (-1)^{s-1}\ffrac{(\q - \q^{-1})^{2m}}{([m]!)^2}\,
  \Bigl(\sum_{n=0}^{s-1}q^{-m(m+s-2n)-(s-1-2n)}\Bnm^+_{n,m}(s)K^{m+s-1-2n}\\
  + \sum_{k=0}^{p-s-1}q^{-m(m-s-2k)+s+1+2k}\Bnm^-_{k,m}(p-s)K^{m-s-1-2k}
  \Bigr)E^m F^m,
\end{multline}
with $\alpha_s$ given in~\eqref{sigma-def3}, and the
coefficients $\Bnm^+_{n,m}(s)$ and $\Bnm^-_{k,m}(p-s)$ are given
in~\eqref{Bnm-pm}, \eqref{Bnm-plus-2}, and \eqref{Bnm-min-2}.
\end{itemize}

\subsection{The modular group action on the center}\label{subsec:modular-qg}
The $\SLiiZ$-action on the $\UresSL2$ center $\cZ$ is defined as~\cite{[FGST],[FGST3]}
\begin{equation}\label{theST}
  \begin{split}
    \modS:{}& a\mapsto \Radford\bigl(\Drinfeld^{-1}(a)\bigr),\\
    \modT:{}& a\mapsto\AT\,\modS(\ribbon\modS^{-1}(a)),
  \end{split}
\end{equation}
which follows~\cite{[Lyu],[LM],[Kerler]}.  Here, the Drinfeld map
$\Drinfeld$ and the Radford map $\Radford$ are defined
in~\eqref{drinfeld-def} and~\eqref{radford-def}, respectively,
the central charge $c$
of $(1,p)$ models is given in~\eqref{centr-charge}, and the ribbon
 central element $\ribbon = e^{2\rmi\pi L_0}$, where $L_0$ is the zero
 mode of the energy-momentum tensor for the $(1,p)$ models
 (see~\cite{[FGST],[FGST3]}).

  There are two bases in $\cZ$. The first one is the Radford basis
\begin{equation}\label{basis-Rad-ind1}
\{\Radford^\pm(s),\;\Radford(r)| 1\leq s\leq p, 1\leq r\leq p-1\}
\end{equation}
evaluated in~\eqref{phi-explicit} and~\eqref{Rad-PBW}, and the second
one is the Drinfeld basis 
\begin{equation}\label{basis-Dr-ind1}
\{\Drinfeld^\pm(s),\;\Drinfeld(r)| 1\leq s\leq p, 1\leq r\leq p-1\}
\end{equation}
evaluated in~\eqref{the-cchi} and~\eqref{the-pscchi}. These bases are
related by the $\modS$-transformation,
\begin{equation}\label{mod-center-S}
      \modS(\Drinfeld^\pm(r))=\Radford^\pm(r), \quad 
      \modS(\Drinfeld(r))=\Radford(r).
\end{equation}

\begin{Thm}\label{thm:center-mult}
The only nonzero multiplications in $\ZZ$ with respect to
  \begin{itemize}
  \item
    the Drinfeld-basis~\eqref{basis-Dr-ind1} are given by \footnote{The
    multiplication~\eqref{the-fusion} was calculated in~\cite{[FGST]}
    and was first obtained in~\cite{[FHST]} in the $(1,p)$ models
    Verlinde-formula context.}, for $1\leq r,s \leq p$, and $\alpha,\beta=\pm$,
\begin{equation}\label{the-fusion}
  \Drinfeld^{\alpha}(r)\Drinfeld^{\beta}(s)
  =\sum_{\substack{t=|r - s| + 1\\
      \mathrm{step}=2}}^{r + s - 1}
  \tilde\Drinfeld^{\alpha\beta}(t),
\end{equation}
where
\begin{equation*}
  \tilde\Drinfeld^{\alpha}(t)
  =
  \begin{cases}
    \Drinfeld^{\alpha}(t),&1\leq t\leq p,\\
    \Drinfeld^{\alpha}( 2p - t) + 2\Drinfeld^{-\alpha}(t - p), & p + 1 \leq t
    \leq 2p - 1,
  \end{cases}
\end{equation*}
and, for $1\leq r,s \leq p-1$,
\begin{align}
& \Drinfeld^+(r)\Drinfeld(s) = \sum_{\substack{i=|r-s|+1\\
  \mathrm{step}=2}}^{\mathrm{min}(r+s-1,2p-r-s-1)} \Drinfeld(i),\label{mult-Dr+-NDr}\\
& \Drinfeld^-(p-r)\Drinfeld(s) = -\!\!\!\sum_{\substack{i=|r-s|+1\\
  \mathrm{step}=2}}^{\mathrm{min}(r+s-1,2p-r-s-1)} \Drinfeld(i).\label{mult-Dr--NDr}
\end{align}
  \item
the Radford-basis~\eqref{basis-Rad-ind1} are given
    by, for $1\leq r,s\leq p-1$,
\begin{align}
&\Radford(r)\,\Radford(r) = \; \ffrac{1}{\Svac_{[r;1]}} \bigl(
 \Radford(r) - \ffrac{\Svac_{[r;2]}}{\Svac_{[r;1]}}\Radford^{+}(r) 
-\ffrac{\Svac_{[r;3]}}{\Svac_{[r;1]}}\Radford^{-}(p-r)\bigr),\label{mult-Rad}\\
&\Radford(r)\,\Radford^+(r) = \; \ffrac{1}{\Svac_{[r;1]}}\, \Radford^+(r),\quad
\Radford(r)\,\Radford^-(p-r) = \; \ffrac{1}{\Svac_{[r;1]}}\, \Radford^-(p-r),\label{mult-Rad-pm}\\
&\Radford^{+}(p)\,\Radford^{+}(p) = \; \ffrac{1}{\Svac_{[p;1]}}\, \Radford^{+}(p),\quad
\Radford^{-}(p)\,\Radford^{-}(p) = \; \ffrac{1}{\Svac_{[p+1;1]}}\,
\Radford^{-}(p).\label{mult-Rad-pp}
\end{align}
Here, $(\Svac_{[r;\beta]})=(S_{[1;2][r;\beta]})$, $1\leq r\leq p+1,
1\leq \beta\leq n_r$, is the ``vacuum'' row, which corresponds to the
$\modS$-transformation of the unity $\Drinfeld^+(1)=\one$, and
$S_{[1;2][r;\beta]}$ are defined in~\eqref{S-3x3} and~\eqref{S-3x2}.
  \end{itemize}
\end{Thm}

The proof of Thm.~\bref{thm:center-mult} is given in~\bref{sec:proof-Thm}.

Next, we recall that the space $\ZZcft$ of torus amplitudes and the
center $\ZZ$ of the quantum group are identified as modular group
representations~\cite{[FGST]} and are isomorphic as associative
commutative algebras. Therefore, identifying the elements in $\ZZcft$
and $\ZZ$ with the same modular-transformation properties
\begin{gather*}
\Drinfeld^{\pm}(s)\mapsto \chi^{\pm}_s(\tau),\qquad
\Drinfeld(s)\mapsto \chi_s(\tau),\\
\Radford^{\pm}(s)\mapsto \chi^{\pm}_s(-1/\tau),\qquad
\Radford(s)\mapsto \chi_s(-1/\tau),
\end{gather*}
we thus get two special bases in $\ZZcft$. The first one is the basis
of the characters and pseudocharacters given in~\eqref{char-def}
and~\eqref{pseudochar-def}, respectively, which corresponds to the
Drinfeld basis in $\ZZ$. The second one is the
basis~\eqref{basis-phi}, which corresponds to the Radford basis in
$\ZZ$. We thus obtain Prop.~\bref{prop-intro:chi-mult} using
~\eqref{the-fusion}-\eqref{mult-Dr--NDr}, and
Prop.~\bref{prop-intro:phi-mult}
using~\eqref{mult-Rad}-\eqref{mult-Rad-pp}.

\begin{rem}
The structure constants in~\eqref{mult-Dr+-NDr} coincide with the
structure constants in $\hSL2_k$ fusion algebra at the level
$k\,{=}\,p\,{-}\,2$. We also note that the multiplications similar to
the ones in
\eqref{mult-Dr+-NDr} and~\eqref{mult-Dr--NDr} appear
  in~\cite{[Flohr:2007]} but for a different basis in the
  space of vacuum torus amplitudes.
\end{rem}

\section{Fusion rules and a generalized Verlinde-formula}\label{sec:gen-Verlinde}
Here, we first give notes toward generalization of the classical
Ver\-lin\-de-formula~\cite{[Verlinde]} to nonsemisimple cases. Then,
we use the results in~\bref{thm:center-mult} to obtain a generalized
Verlinde-formula for the $(1,p)$ models.
\subsection{Toward generalization of the Verlinde formula}\label{sec:gen-verl}
We assume that the $S$-trans\-for\-ma\-ti\-on from the modular group acting on
a finite-dimensional associative algebra, which we denote as $\ZZcft$, satisfy
\begin{equation*}
S^2|_{\ZZcft}=\text{id}\,,
\end{equation*}
and block-diagonalizes the structure constants
(the ``fusion'' coefficients) in 
$\ZZcft$ with respect to a distinguished basis $\{\Drinfeld_i| i\in
I\}$, where $I$ is some finite set, to the block-diagonal structure
\begin{equation}\label{block-str-gen}
n_1\times n_1 \oplus \dots \oplus n_r\times n_r \oplus \dots \oplus
n_P\times n_P,
\end{equation}
where $P$ is the number of Jordan blocks, $1\leq r\leq P$, and
$n_r \geq 1$ is the rank of the $r$-th Jordan block. We also assume that
we have the unity $\one$ in $\ZZcft$.

We group the distinguished basis elements $\{\Drinfeld_i| i\in I\}$ in
$\ZZcft$ into blocks (sets) with respect to the block-diagonal
structure~\eqref{block-str-gen},
\begin{equation*}
\{\Drinfeld_{[r;\alpha]}\equiv \Drinfeld_{n_1+\dots+n_{r-1}+\alpha}| 1
\leq r\leq P, 1\leq \alpha\leq n_r\}.
\end{equation*}
 We also arrange the structure constants, describing the
 multiplication between $\Drinfeld_{[r;\alpha]}$ and
 $\Drinfeld_{[s;\beta]}$ with the result in $[k;\gamma]$-block, 
into a rank-$3$ tensor $N_{[r;\alpha][s;\beta]}^{[k;\gamma]}$,
\begin{equation}\label{Dr-struct-const}
\Drinfeld_{[r;\alpha]}\Drinfeld_{[s;\beta]} =
\sum_{k=1}^{P}\sum_{\gamma=1}^{n_k}
N_{[r;\alpha][s;\beta]}^{[k;\gamma]}\Drinfeld_{[k;\gamma]}.
\end{equation}

For $1\leq r\leq P$ and $1\leq \alpha\leq n_r$, we denote
\begin{equation*}
\Radford_{[r;\alpha]} = S(\Drinfeld_{[r;\alpha]}) =
 \sum_{j=1}^{P}\sum_{\beta=1}^{n_j}
    S_{[r;\alpha][j;\beta]} \Drinfeld_{[j;\beta]},
\end{equation*}
 and the structure constants in this basis by
$\Radford_{[r;\lambda][s;\mu]}^{[k;\nu]}$,
\begin{equation}\label{eq:rad-rad-gen}
\Radford_{[r;\lambda]}\Radford_{[s;\mu]} = \sum_{k=1}^{P} \sum_{\nu=1}^{n_k}
\Radford_{[r;\lambda][s;\mu]}^{[k;\nu]} \Radford_{[k;\nu]}.
\end{equation}

\begin{thm}\label{thm:Verlinde-gen} The structure constants
  $N_{[r;\alpha][s;\beta]}^{[k;\gamma]}$ in $\ZZcft$ are reproduced
  from the $S$-matrix action,
\begin{equation}\label{fusion-coeff-gen}
N_{[r;\alpha][s;\beta]}^{[k;\gamma]} =
\sum_{l=1}^{P}\sum_{\lambda, \mu, \nu
=1}^{n_l}S_{[r;\alpha][l;\lambda]}
S_{[s;\beta][l;\mu]}\Radford_{[l;\lambda][l;\mu]}^{[l;\nu]}
S_{[l;\nu][k;\gamma]},
\end{equation}
where the structure constants
$\Radford_{[l;\lambda][l;\mu]}^{[l;\nu]}$ are
solutions of the following equations, for $1\leq l\leq P$ and $1\leq \mu,\nu\leq n_l$,
\begin{equation}\label{str-rad-rad-sol}
\sum_{\lambda=1}^{n_l}
\Svac_{[l;\lambda]} \Radford_{[l;\lambda][l;\mu]}^{[l;\nu]} = \delta_{\mu,\nu}.
\end{equation}
Here, $(\Svac_{[l;\lambda]})=(S_{[1;2][l;\lambda]})$, $1\leq l\leq P$,
$1\leq \lambda\leq n_l$, is the ``vacuum'' row, which corresponds to
the $S$-transformation of the unity $\Drinfeld_2=\one$.
\end{thm}
\begin{proof}
The formula~\eqref{fusion-coeff-gen} is just relation between
structure constants in different bases related by the
$S$-transformation.

Let us consider the $S$-transformation of the unity. We have
\begin{equation*}
S(\one) = \sum_{i=1}^{P}\sum_{\lambda=1}^{n_i} \Svac_{[i;\lambda]} \Drinfeld_{[i;\lambda]}.
\end{equation*}
Next, we recall the assumption $S^{2}|_{\ZZcft}=\text{id}$. Therefore, we have
\begin{equation*}
\one = \sum_{i=1}^{P}\sum_{\lambda=1}^{n_i} \Svac_{[i;\lambda]} \Radford_{[i;\lambda]}.
\end{equation*}
Hence, using~\eqref{eq:rad-rad-gen} and the assumption that the $S$-transformation
block-diagonalizes the structure constants in $\ZZcft$ to the
block-diagonal structure~\eqref{block-str-gen}, we obtain the identities
\begin{gather*}
\Radford_{[l;\mu]} = \sum_{i=1}^{P}\sum_{\lambda=1}^{n_i}
\Svac_{[i;\lambda]} \Radford_{[i;\lambda]} \Radford_{[l;\mu]} = \sum_{\lambda, \nu=1}^{n_l}
\Svac_{[l;\lambda]} \Radford_{[l;\lambda][l;\mu]}^{[l;\nu]}
\Radford_{[l;\nu]},
\end{gather*}
for $1\leq l\leq P$ and $1\leq \mu\leq n_l$. These identities give
the equations~\eqref{str-rad-rad-sol} on the structure constants
$\Radford_{[l;\lambda][l;\mu]}^{[l;\nu]}$ in~\eqref{eq:rad-rad-gen}. This
completes the proof.
\end{proof}

\begin{rem}
The Verlinde-like formula~\eqref{fusion-coeff-gen} reproduces the classical
Verlinde-for\-mu\-la~\cite{[Verlinde]} for semisimple (rational) theories,
in which the $S$-transformation diagonalizes the structure constants in
a fusion algebra. Indeed, the equations~\eqref{str-rad-rad-sol} take
the following form
\begin{equation*}
\Svac_{[l;1]} \Radford_{[l;1][l;1]}^{[l;1]} = 1,
\end{equation*}
and we get the Verlinde formula
\begin{equation*}
N_{rs}^{k} \equiv N_{[r;1][s;1]}^{[k;1]} =
\sum_{l=1}^{P}\ffrac{S_{rl}S_{sl}S_{lk}}
{\Svac_{l}}.
\end{equation*}
\end{rem}

\subsection{A generalized Verlinde-formula for the $(1,p)$ models}
\label{sec:fus-1p}
Here, we apply the results in~\bref{thm:Verlinde-gen}
and~\bref{thm:center-mult} to obtain the structure constants
(``generalized'' fusion coefficients) with respect to the basis of the
characters and pseudocharacters in $\ZZcft$ from the modular-group
action on $\ZZcft$.
These structure constants coincide with the structure constants in the
Drinfeld basis~\eqref{basis-Dr-ind1},
\begin{equation}\label{eq:str-const-Dr-def}
\Drinfeld_{[r;\alpha]}\Drinfeld_{[s;\beta]} =
\sum_{k=1}^{p+1}\sum_{\gamma=1}^{n_k}
N_{[r;\alpha][s;\beta]}^{[k;\gamma]}\Drinfeld_{[k;\gamma]},
\end{equation}
where we write the Drinfeld basis \eqref{basis-Dr-ind1} in the
 following $2$-index notation parallel to the one
 in~\eqref{basis-chi-def},
\begin{multline}\label{basis-Dr}
\{\Drinfeld_{[s;\alpha]} | 1\leq s\leq p+1, 1\leq \alpha \leq n_s\}=\\
=  \{\Drinfeld(1),\,\Drinfeld^{+}(1),\, \Drinfeld^{-}(p-1),\,
\dots,\, \Drinfeld(p-1),\, \Drinfeld^{+}(p-1),\,\Drinfeld^{-}(1),\,
\Drinfeld^{+}(p),\, \Drinfeld^{-}(p)\},
\end{multline}
where $n_s=3$, for $1\leq s\leq p-1$, and $n_s=1$, for $s=p,p+1$, that is,
 we group the Drinfeld-basis elements into $p-1$ triplets
 $\{\Drinfeld(s),\Drinfeld^+(s),\Drinfeld^-(p-s)\}$, for $1\leq
 s\leq p-1$, and into two singlets $\Drinfeld^+(p)$,
 $\Drinfeld^-(p)$.  Similarly, for the Radford basis
 \eqref{basis-Rad-ind1}, we denote
\begin{multline}\label{basis-Rad}
 \{\Radford_{[s;\alpha]} | 1\leq s\leq p+1, 1\leq \alpha \leq n_s\} =\\
 = \{\Radford(1),\,\Radford^{+}(1),\, \Radford^{-}(p-1),\, 
\dots,\,\Radford(p-1),\,\Radford^{+}(p-1),\,\Radford^{-}(1),\,
\Radford^{+}(p),\, \Radford^{-}(p)\}.
\end{multline}

 In~\cite{[FGST]}, it was shown that the $\SLiiZ$ action on the center $\ZZ$
is equivalent to the one on the space $\ZZcft$ of torus amplitudes for the
$(1,p)$ models. Therefore, elements of the basis \eqref{basis-Dr} can
be linearly expressed with respect to the basis \eqref{basis-Rad},
  \begin{equation}\label{Dr-S-Rad}
    \Drinfeld_{[s;\alpha]}=\sum_{j=1}^{p+1}\sum_{\beta=1}^{n_j}
    S_{[s;\alpha][j;\beta]} \Radford_{[j;\beta]}
  \end{equation}
 and vice versa 
  \begin{equation*}
    \Radford_{[s;\alpha]}=\sum_{j=1}^{p+1}\sum_{\beta=1}^{n_j}
    S_{[s;\alpha][j;\beta]} \Drinfeld_{[j;\beta]},
  \end{equation*}
where the $S$-matrix elements  $S_{[s;\alpha][j;\beta]}$ are given
in~\eqref{S-block}-\eqref{S-2x2}.

Therefore, the structure constants
$N_{[r;\alpha][s;\beta]}^{[k;\gamma]}$ in~\eqref{eq:str-const-Dr-def}
can be calculated by use~\eqref{fusion-coeff-gen}
and~\eqref{str-rad-rad-sol}, where we must set $P=p+1$, $n_r=3$, for
$1\leq r\leq p-1$, and $n_p=n_{p+1}=1$; and the structure constants
$\Radford_{[l;\lambda][l;\mu]}^{[l;\nu]}$ in the Radford basis
\eqref{basis-Rad} are solutions of~\eqref{str-rad-rad-sol} and
coincide with~\eqref{mult-Rad}-\eqref{mult-Rad-pp}. For $1\leq l \leq
p-1$, these structure constants are
\begin{equation}\label{str-const-Rad-1}
\Radford_{[l;1][l;\mu]}^{[l;\nu]} =
\ffrac{1}{\Svac_{[l;1]}}\bigl(\sum_{i=1}^3\delta_{\mu,i}\delta_{\nu,i}
- \ffrac{\Svac_{[l;2]}}{\Svac_{[l;1]}}\delta_{\mu,1}\delta_{\nu,2} -
\ffrac{\Svac_{[l;3]}}{\Svac_{[l;1]}}\delta_{\mu,1}\delta_{\nu,3}\bigr),
\end{equation}
and
\begin{equation}\label{str-const-Rad-2}
\Radford_{[l;2][l;\mu]}^{[l;\nu]} =
\ffrac{1}{\Svac_{[l;1]}}\delta_{\mu,1}\delta_{\nu,2},\qquad
\Radford_{[l;3][l;\mu]}^{[l;\nu]}=
\ffrac{1}{\Svac_{[l;1]}}\delta_{\mu,1}\delta_{\nu,3},
\end{equation}
and, for $l=p,p+1$, 
$\Radford_{[l;1][l;\mu]}^{[l;\nu]} =
\ffrac{1}{\Svac_{[l;1]}}\delta_{\mu,1}\delta_{\nu,1}$. 
Hence, using~\eqref{fusion-coeff-gen}, we finally get 
\begin{equation}\label{fusion-coef-big}
N_{[r;\alpha][s;\beta]}^{[k;\gamma]}
=\sum_{l=1}^{p+1}\sum_{\lambda=1}^{n_l}
\ffrac{\Svac_{[l;1]} S^{[r;\alpha]}_{[l;1]}
S^{[s;\beta]}_{[l;\lambda]} +
\Svac_{[l;1]}S^{[r;\alpha]}_{[l;\lambda]}
S^{[s;\beta]}_{[l;1]}-\Svac_{[l;\lambda]}S^{[r;\alpha]}_{[l;1]}
S^{[s;\beta]}_{[l;1]}}{\bigl(\Svac_{[l;1]}\bigr)^2}S^{[l;\lambda]}_{[k;\gamma]},
\end{equation}
where we set $S^{[r;\alpha]}_{[l;\lambda]}\equiv
S_{[r;\alpha][l;\lambda]}$. The generalized
Verlinde-formula~\eqref{fusion-coef-big} reproduces the structure
constants in $\ZZcft$ (given in~\bref{prop-intro:chi-mult}) with
respect to the basis~\eqref{basis-chi-def} of the characters and pseudocharacters.
\begin{rem}\label{rem:fusion-Verl}
The structure constants~\eqref{fusion-coef-big}, for
$\alpha,\beta,\gamma=2,3$, coincide with the fusion coefficients of
the $(1,p)$ models obtained in~\cite{[FHST]} and give the structure
constants in the Grothendieck ring for $\UresSL2$~\cite{[FGST]}. At
the same time, the right-hand side of~\eqref{fusion-coef-big}, for
$\alpha,\beta,\gamma=2,3$, seems to be related to the formula $(5.28)$
from~\cite{[GR2]} giving the same fusion coefficients.
\end{rem}

\section{The space of boundary states and the center of the quantum
  group} \label{sec:boundary-st} Here, we analyze boundary states in
 the $(1,p)$ models.  In~\bref{sec:irbs}, we choose a basis in the
 space of boundary states in such a way that this basis correspond to
 the Radford basis in the quantum group center
 $\ZZ$. In~\bref{sec:cdbs}, we show that the states that correspond to
 the Drinfeld elements in $\ZZ$ satisfy the Cardy conditions and
 therefore are Cardy boundary states.

\subsection{Ishibashi and Radford boundary states\label{sec:irbs}}
The equivalence between the triplet algebra $\algW_p$- and
$\UresSL2$-representaions categories leads to an isomorphism $\ZisoZ$
between the center $\ZZ$ of $\UresSL2$ and the space of boundary
states in the $(1,p)$ models.  To describe the isomorphism $\ZisoZ$,
we first note that the space of boundary states can be identified with
the space $\ZZcft$ of vacuum torus amplitudes in the following way.
We let $\iket{\pm,s}$ with $1\leq s\leq p$ and $\iket{s}$ with $1\leq
s\leq p-1$ denote the Ishibashi states satisfying (cf.~\cite{[GR2]})
\begin{align}
   &\ibra{r}q^{\half(L_0+\bar L_0-\frac{c}{12})}\iket{s} = \delta_{r,s}
  \ffrac{1}{\Svac_{[s;1]}} \bigl(\chi_s(q)
  - \ffrac{\Svac_{[s;2]}}{\Svac_{[s;1]}}\chi^+_{s}(q) -
  \ffrac{\Svac_{[s;3]}}{\Svac_{[s;1]}} \chi^-_{p-s}(q)\bigr),\label{ishibash-rs}\\
  &\ibra{r}q^{\half(L_0+\bar L_0-\frac{c}{12})} \iket{+,s} = 
  \delta_{r,s}\ffrac{1}{\Svac_{[s;1]}} \chi^+_s(q),\\
  &\ibra{r}q^{\half(L_0+\bar L_0-\frac{c}{12})} \iket{-,p-s} = 
  \delta_{r,s}\ffrac{1}{\Svac_{[s;1]}} \chi^-_{p-s}(q),\\
  &\ibra{\alpha,r}q^{\half(L_0+\bar L_0-\frac{c}{12})}\iket{\beta,s}=0,
  \qquad\alpha,\beta=\pm,\quad 1\leq r\leq p-1,\quad 1\leq s\leq p,\\
  &\ibra{+,p}q^{\half(L_0+\bar
  L_0-\frac{c}{12})}\iket{+,p}=\ffrac{1}{\Svac_{[p;1]}}\chi^+_p(q),\\
  &\ibra{-,p}q^{\half(L_0+\bar L_0-\frac{c}{12})}\iket{-,p}=\ffrac{1}{\Svac_{[p+1;1]}}\chi^-_p(q),\\
  &\ibra{\pm,p}q^{\half(L_0+\bar L_0-\frac{c}{12})}\iket{\mp,p}=0,\label{ishibash-pp=mp}
\end{align}
where the characters $\chi^{\pm}_s(q)$ are given
in~\eqref{eq:characters}, and the
pseudocharacters $\chi_s(q)$ 
in~\eqref{eq:pseudochar}.
\begin{rem}
We note that the Ishibashi states proposed in~\cite{[GR2]} are in
correspondence with those introduced
in~\eqref{ishibash-rs}-\eqref{ishibash-pp=mp},
\begin{gather*}
\iket{P_t} = \Svac_{[t;1]}\iket{t} + \Svac_{[t;2]}\iket{+,t} +
\Svac_{[t;3]}\iket{-,p-t},\\
\iket{U_t} = \iket{+,t} + \iket{-,p-t},\qquad
\iket{U_p^{+}} = \Svac_{[p;1]}\iket{+,p},\quad
\iket{U_p^{-}} = \Svac_{[p+1;1]}\iket{-,p},
\end{gather*}
where we note that $\Svac_{[t;2]}=\Svac_{[t;3]}$. In what follows, we
need all $(3p-1)$ Ishibashi states but not only $2p$ of them as in~\cite{[GR2]}.
\end{rem}

We next define the isomorphism $\ZisoZ$ between the center $\ZZ$ and
the space of boundary states, which we similarly denote as $\ZZcft$,
\begin{equation*}
  \ZisoZ:\ZZ\to\ZZcft
\end{equation*}
by the formula
\begin{equation}\label{isoF-def}
  \ZisoZ(\Radford^{\pm}(s))=\iket{\pm,s},\qquad\ZisoZ(\Radford(s))=\iket{s}.
\end{equation}
We call the states $\iket{\pm,s}$, and $\iket{s}$ \textit{the
Radford boundary states}. The
overlaps~\eqref{ishibash-rs}-\eqref{ishibash-pp=mp} between
the Radford boundary states can be written shortly in the $2$-index
notation parallel to the one in~\eqref{basis-chi-def},
\begin{equation}\label{rad-bnd-states}
\ibra{[r;\alpha]}q^{\half(L_0+\bar L_0-\frac{c}{12})}\iket{[s;\beta]} =
  \sum_{l=1}^{p+1}\sum_{\gamma=1}^{n_l}
  \Radford_{[r;\alpha][s;\beta]}^{[l;\gamma]}\chi_{[l;\gamma]}(q)
= \delta_{r,s}\sum_{\gamma=1}^{n_r}
  \Radford_{[r;\alpha][r;\beta]}^{[r;\gamma]}\chi_{[r;\gamma]}(q),
\end{equation}
where $n_r=3$, for $1\leq r\leq p-1$, and $n_r=1$, for
$r=p,p+1$, and we set
\begin{gather}\label{basis-ishib-def}
\bigl(\iket{[s;1]},\iket{[s;2]},\iket{[s;3]}\bigr) =
  (\iket{s},\iket{+,s},\iket{-,p-s}),\qquad 1\leq s\leq p-1,\\
\iket{[p;1]} = \iket{+,p},\quad \iket{[p+1;1]} = \iket{-,p},\notag
\end{gather}
and analogously for the bra-vectors $\ibra{[r;\alpha]}$;
$\chi_{[r;\gamma]}(q)$ are defined in~\eqref{basis-chi-def}, and the
structure constants $\Radford_{[r;\alpha][r;\beta]}^{[r;\gamma]}$ are
given in~\eqref{str-const-Rad-1} and~\eqref{str-const-Rad-2}.

\subsection{Cardy and Drinfeld boundary states} \label{sec:cdbs}
 We introduce Drinfeld boundary states as the $\ZisoZ$ images of the
 Drinfeld elements~\eqref{basis-Dr-ind1},
\begin{equation*}
  \ZisoZ(\Drinfeld^{\pm}(s))=\cket{\pm,s},\qquad\ZisoZ(\Drinfeld(s))=\cket{s}.
\end{equation*}
We call the states
$\cket{\pm,s}$ and $\cket{s}$ \textit{the Drinfeld boundary
states}. From~\eqref{isoF-def} and~\eqref{Dr-S-Rad}, the Drinfeld boundary
states are linearly expressed
  with respect to the Radford boundary states as
\begin{equation}\label{Drinf-bound-st}
    \cket{[s;\alpha]}=\sum_{j=1}^{p+1}\sum_{\beta=1}^{n_j}
    S_{[s;\alpha][j;\beta]}  \iket{[j;\beta]},
  \end{equation}
where we also use the $2$-index notation parallel to the one
introduced in~\eqref{basis-ishib-def}.
\begin{prop}
The Drinfeld boundary states~\eqref{Drinf-bound-st} satisfy the Cardy
condition
 \begin{equation}\label{cardy-cond}
  \cbra{[r;\alpha]}q^{\half(L_0+\bar
  L_0-\frac{c}{12})}\cket{[s;\beta]}=\sum_{k=1}^{p+1}
  \sum_{\gamma=1}^{n_k} N_{[r;\alpha][s;\beta]}^{[k;\gamma]}
  \chi_{[k;\gamma]}(\tilde q),
\end{equation}
where $\tilde q = e^{-2\rmi\pi/\tau}$, and the structure
 constants $N_{[r;\alpha][s;\beta]}^{[k;\gamma]}$ are given
 in~\eqref{fusion-coef-big}.
\end{prop}
\begin{proof}
This trivially follows from~\eqref{Drinf-bound-st},
\eqref{rad-bnd-states}, and~\eqref{fusion-coeff-gen}.
\end{proof}
The $\ZisoZ$ images of the Drinfeld elements in the quantum group
center satisfy the Cardy condition~\eqref{cardy-cond} and are
therefore the Cardy states.
\begin{rem}
We have $3p-1$ Drinfeld boundary states that formally satisfy the
Cardy condition~\eqref{cardy-cond} but only $2p$ of them,
$\cket{\pm,r}$, for $1\leq r\leq p$, have transparent physical meaning
(see the last paragraph in Sec.~5.1~\cite{[GR2]}). We note also that
the Cardy states $\cket{\pm,r}$ coincide with $\cket{(r,\pm)}$ from~\cite{[GR2]}.
\end{rem}

\section{Conclusions}\label{sec:concl}
In this paper we propose a constructive method to study the boundary
theories. The method is based on the well-known Kazhdan--Lusztig
correspondence stated for the $(1,p)$ models in~\cite{[FGST]}, and for
$(p,q)$-logarithmic models in~\cite{[FGST3],[FGST4]}. The
Kazhdan--Lusztig correspondence for the $(1,p)$ models is an
equivalence~\cite{[FGST2]} between representation categories of the
triplet algebra $\algW_p$ and the \textit{restricted} quantum group
$\UresSL2$, with $\q=e^{\frac{\rmi\pi}{p}}$. The equivalence leads to an
isomorphism between the space of boundary states in the $(1,p)$ models
and the center $\ZZ$ of $\UresSL2$.

We found a basis in the quantum-group center $\ZZ$ in which the
structure constants are integer numbers and $2p$ elements of the basis
are Drinfeld images of irreducible module characters, which span the
Grothendieck ring. Under identification of the quantum group center
and the space of $(1,p)$ model boundary states, the elements of the
Drinfeld basis are mapped to the states satisfying the Cardy
condition. Thus, we have $3p-1$ such states, but only $2p$ of them
have transparent physical meaning and $p-1$ satisfy the Cardy
condition only formally because negative structure constants can not
be interpreted as multiplicities of any representations. These
findings raise a good question about physical meaning of these
additional $p-1$ Cardy states.

It is interesting to find the Radford and Drinfeld boundary states in
terms of a free scalar field, which is used for formulation of the
$(1,p)$ models as screening kernels. It would allow us to understand
better a meaning of $p-1$ boundary states $\cket{[r;1]}$.

We also propose the generalized
Verlinde-formula~\eqref{fusion-coef-big}, which gives the integer
structure constants in the whole $(3p-1)$-dimensional space of vacuum
torus amplitudes for the $(1,p)$ models, in which the fusion algebra
is a $2p$-dimensional subalgebra (cf.~\cite{[Flohr:2007]}) This
formula can therefore be considered as a generalization of $(1,p)$
model Verlinde formulas derived in~\cite{[FHST],[Flohr:2007],[GR2]}.

We hope that our reslults can be extended into the logarithmic $(p,q)$
models~\cite{[FGST4]}, for which the Kazhdan--Lusztig dual
quantum-group is proposed in~\cite{[FGST3]}. We thus conjecture that
there are $\half(3p-1)(3q-1)$ Cardy states in these $(p,q)$ models. It
would be also interesting to obtain a generalized Verlinde-formula for
the logarithmic $(p,q)$ models and compare results with the ones
in~\cite{[Semi],[EF]}.

\subsubsection*{Acknowledgments} We are grateful to Alexei Semikhatov
and Boris Feigin for useful comments. AMG is also grateful to George
Mutafyan for useful discussions. A part of the paper was written
during our stay in Kyoto University, and we are grateful to T.~Miwa
for the kind hospitality extended to us.  This paper was supported in
part by the RFBR Grant~07-01-00523 and by the Grant~LSS-1615.2008.2.
The work of AMG was also supported by the ``Landau'', ``Dynasty'', and
``Science Support'' foundations. The work of IYuT was also supported
by RFBR Grant~05-02-17217 and the ``Dynasty'' foundation.

\appendix

\section{$W$ characters and pseudocharacters}
The $\algW_p$ characters are given by~\cite{[FHST],[F-95]}
  \begin{equation}\label{eq:characters}
    \begin{aligned}
      \chi^{+}_{s}(q)&=\mfrac{1}{\eta(q)}
      \Bigl(\ffrac{s}{p}\,\theta_{p{-}s,p}(q)
      + 2\,\theta'_{p{-}s,p}(q)\Bigr),\\[4pt]
      \chi^{-}_{s}(q)&=
      \mfrac{1}{\eta(q)}
      \Bigl(\ffrac{s}{p}\,\theta_{s,p}(q) - 2\,\theta'_{s,p}(q)\Bigr),
    \end{aligned}\qquad
    1\leq s\leq p,
  \end{equation}
and the $\algW_p$ pseudocharacters are
\begin{equation}\label{eq:pseudochar}
\chi_{s\andp}(q) =
\mfrac{2a_0}{\eta(q)}\log(q)\theta'_{p{-}s,p}(q), \qquad 1\leq s\leq p-1
\end{equation}
\noindent
(see also~\cite{[FG]}). Here, we use the eta function
\begin{align*}
  \eta(q)&=q^{\frac{1}{24}} \prod_{n=1}^{\infty} (1-q^n)
  \\
  \intertext{and the theta functions} 
  \theta_{s,p}(q,z)&=\sum_{j\in\oZ + \frac{s}{2p}} q^{p j^2} z^j,
  \quad |q|<1,~z\in\oC\,,
\end{align*}
and set $\theta_{s,p}(q)\,{:=}\,\theta_{s,p}(q,1)$ and
$\theta'_{s,p}(q)\,{:=}\,z\frac{\dd}{\dd
  z}\theta_{s,p}(q,z)\!\!\bigm|_{z=1}$.

\section{Irreducible and projective $\UresSL2$-modules}\label{app:irr-proj}
Here, we recall the definition of irreducible and projective
$\UresSL2$-modules~\cite{[FGST]}.
\subsection{Irreducible $\UresSL2$-modules}
The irreducible $\UresSL2$-modules are labeled by their highest
weights $\q^{s-1}$, where $s\in\oZ/2p\oZ$.  We also parameterize the
same highest weights as $\alpha\q^{s-1}$, where $\alpha\,{=}\,\pm$ and $1\leq
s\leq p$.  Then, for $1\leq s\leq p$, the irreducible module with the
highest weight $\pm \q^{s-1}$ is denoted by $\XX^{\pm}_{s}$.  The
dimension-$s$ module $\XX^{\pm}_{s}$ is spanned by elements
$\stprp_n^{\pm}$, $0\leq n\leq s{-}1$, where $\stprp_0^{\pm}$ is the
highest-weight vector and the left action of the algebra is given~by
\begin{align*}
  K \stprp_n^{\pm} &=
  \pm \q^{s - 1 - 2n} \stprp_n^{\pm},\\
  E \stprp_{n}^{\pm} &=
  \pm [n][s - n]\stprp_{n - 1}^{\pm},\\
  F \stprp_n^{\pm} &= \stprp_{n + 1}^{\pm},
\end{align*}
where we set $\stprp_{-1}^{\pm}=\stprp_{s}^{\pm}=0$.

\subsection{Projective $\UresSL2$-modules}\label{proj-mod-base}
The module $\PP^{\pm}_s$, $1\leq s\leq p-1$, is the
projective module whose irreducible quotient is given
by~$\XX^{\pm}_s$. 
Their structure can be schematically depicted~as
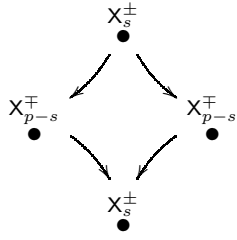
\begin{equation}\label{schem-proj}
  \xymatrix@=10pt{
    &&\stackrel{\XX^{\pm}_s}{\bullet}
    \ar@/^/[dl]
    \ar@/_/[dr]
    &\\
    &\stackrel{\XX^{\mp}_{p - s}}{\bullet}\ar@/^/[dr]
    &
    &\stackrel{\XX^{\mp}_{p - s}}{\bullet}\ar@/_/[dl]
    \\
    &&\stackrel{\XX^{\pm}_s}{\bullet}&
  }
\end{equation}
\subsubsection{$\PP^+_s$}\label{module-L}
Let $s$ be an integer $1\leq s\leq p-1$.  The projective module
$\PP^+_s$ has the basis
\begin{equation}\label{left-proj-basis-plus}
  \{\toppr^{(+,s)}_n,\botpr^{(+,s)}_n\}_{0\le n\le s-1}
  \cup\{\leftpr^{(+,s)}_k,\rightpr^{(+,s)}_k\}_{0\le k\le p-s-1},
\end{equation}
where $\{\toppr^{(+,s)}_n\}_{0\le n\le s-1}$ is the basis
corresponding to the top module in~\eqref{schem-proj},\\
$\{\botpr^{(+,s)}_n\}_{0\le n\le s-1}$ to the bottom ,
$\{\leftpr^{(+,s)}_k\}_{0\le k\le p-s-1}$ to the left, and
$\{\rightpr^{(+,s)}_k\}_{0\le k\le p-s-1}$ to the right module, with
the $\tqalgA$-action given by
\begin{alignat*}{3}
  K\leftpr^{(+,s)}_k&=-\q^{p-s-1-2k}\leftpr^{(+,s)}_k,& \quad
  K\rightpr^{(+,s)}_k&=-\q^{p-s-1-2k}\rightpr^{(+,s)}_k,&
  \quad &0\le k\le p-s-1,\\
  K\botpr^{(+,s)}_n&=\q^{s-1-2n}\botpr^{(+,s)}_n,& \quad
  K\toppr^{(+,s)}_n&=\q^{s-1-2n}\toppr^{(+,s)}_n,& \quad &0\le n\le
  s-1,\\
  E\leftpr^{(+,s)}_k&=-[k][p-s-k]\leftpr^{(+,s)}_{k-1},&
  \quad 0\le k&\le p-s-1
  \quad(\text{with}\quad\leftpr^{(+,s)}_{-1}\equiv0),
  \kern-60pt
\end{alignat*}
\begin{align*}
  E\rightpr^{(+,s)}_k&=
  \begin{cases}
    -[k][p-s-k]\rightpr^{(+,s)}_{k-1}, &1\le k\le p-s-1,\\
    \botpr^{(+,s)}_{s-1}, & k=0,\\
  \end{cases}
  \\
  E\botpr^{(+,s)}_n&=[n][s-n]\botpr^{(+,s)}_{n-1},
  \quad 0\le n\le s-1\quad(\text{with}\quad\botpr^{(+,s)}_{-1}\equiv0),\\
  E\toppr^{(+,s)}_n&=
  \begin{cases}
    [n][s-n]\toppr^{(+,s)}_{n-1}+\botpr^{(+,s)}_{n-1}, &1\le n\le s-1,\\
    \leftpr^{(+,s)}_{p-s-1}, & n=0,\\
  \end{cases}
  \\
  \intertext{and}
  F\leftpr^{(+,s)}_k&=
  \begin{cases}
    \leftpr^{(+,s)}_{k+1}, &0\le k\le p-s-2,\\
    \botpr^{(+,s)}_0, & k=p-s-1,\\
  \end{cases}
  \\
  F\rightpr^{(+,s)}_k&=\rightpr^{(+,s)}_{k+1}, \quad 0\le k\le p-s-1
  \quad(\text{with}\quad\rightpr^{(+,s)}_{p-s}\equiv0),\\
  F\botpr^{(+,s)}_n&=\botpr^{(+,s)}_{n+1}, \quad 0\le n\le s-1
  \quad(\text{with}\quad\botpr^{(+,s)}_s\equiv0),\\
    F\toppr^{(+,s)}_n&=
  \begin{cases}
    \toppr^{(+,s)}_{n+1}, &0\le n\le s-2,\\
    \rightpr^{(+,s)}_0, & n=s-1.
  \end{cases}
\end{align*}

\subsubsection{$\PP^-_{p-s}$}\label{module-P}
Let $s$ be an integer $1\leq s\leq p-1$.  The projective module
$\PP^-_{p-s}$ has the basis
\begin{equation}\label{left-proj-basis-min}
  \{\toppr^{(-,s)}_k,\botpr^{(-,s)}_k\}_{0\le k\le p-s-1}
  \cup\{\leftpr^{(-,s)}_n,\rightpr^{(-,s)}_n\}_{0\le n\le s-1},
\end{equation}
where $\{\toppr^{(-,s)}_k\}_{0\le k\le p-s-1}$ is the basis
corresponding to the top module in~\eqref{schem-proj},\\
$\{\botpr^{(-,s)}_k\}_{0\le k\le p-s-1}$ to the bottom,
$\{\leftpr^{(-,s)}_n\}_{0\le n\le s-1}$ to the left, and
$\{\rightpr^{(-,s)}_n\}_{0\le n\le s-1}$ to the right module, with the
$\tqalgA$-action given by
\begin{alignat*}{3}
  K\botpr^{(-,s)}_k&=-\q^{p-s-1-2k}\botpr^{(-,s)}_k,& \quad
  K\toppr^{(-,s)}_k&=-\q^{p-s-1-2k}\toppr^{(-,s)}_k,&
  \quad &0\le k\le p-s-1,\\
  K\leftpr^{(-,s)}_n&=\q^{s-1-2n}\leftpr^{(-,s)}_n,& \quad
  K\rightpr^{(-,s)}_n&=\q^{s-1-2n}\rightpr^{(-,s)}_n,& \quad &0\le n\le
  s-1,\\
  E\botpr^{(-,s)}_k&=-[k][p-s-k]\botpr^{(-,s)}_{k-1},& \quad
  0\le k&\le p-s-1\quad(\text{with}\quad\botpr^{(-,s)}_{-1}\equiv0),
  \kern-60pt
\end{alignat*}
\begin{align*}
  E\toppr^{(-,s)}_k&=
  \begin{cases}
    -[k][p-s-k]\toppr^{(-,s)}_{k-1}+\botpr^{(-,s)}_{k-1},
    &1\le k\le p-s-1,\\
    \leftpr^{(-,s)}_{s-1}, & k=0,\\
  \end{cases}
  \\
  E\leftpr^{(-,s)}_n&=[n][s-n]\leftpr^{(-,s)}_{n-1},
  \quad 0\le n\le s-1\quad(\text{with}\quad
  \leftpr^{(-,s)}_{-1}\equiv0),\\
  E\rightpr^{(-,s)}_n&=
  \begin{cases}
    [n][s-n]\rightpr^{(-,s)}_{n-1}, &1\le n\le s-1,\\
    \botpr^{(-,s)}_{p-s-1}, & n=0,\\
  \end{cases}
  \\
  \intertext{and}
  F\botpr^{(-,s)}_k&=\botpr^{(-,s)}_{k+1}, \quad 0\le k\le p-s-1
  \quad(\text{with}\quad\botpr^{(-,s)}_{p-s}\equiv0),\\
  F\toppr^{(-,s)}_k&=
  \begin{cases}
    \toppr^{(-,s)}_{k+1}, &0\le k\le p-s-2,\\
    \rightpr^{(-,s)}_0, & k=p-s-1,\\
  \end{cases}
  \\
  F\leftpr^{(-,s)}_n&=
  \begin{cases}
    \leftpr^{(-,s)}_{n+1}, &0\le n\le s-2,\\
    \botpr^{(-,s)}_0, & n=s-1,
  \end{cases}
  \\
  F\rightpr^{(-,s)}_n&=\rightpr^{(-,s)}_{n+1}, \quad 0\le n\le s-1
  \quad(\text{with}\quad\rightpr^{(-,s)}_s\equiv0).
\end{align*}

\section{Radford images of pseudotraces and
 the proof of Thm.~\bref{thm:center-mult}}
Here, we present bulky calculation of Radford images of pseudotraces
and give the proof of Thm.~\bref{thm:center-mult} in \bref{sec:proof-Thm}.

\subsection{PBW-basis action on projectives}
The elements of the PBW-basis of $\UresSL2$ are enumerated as
$E^i\,K^j\,F^\ell$ with $0\leq i\leq p-1$, $0\leq j\leq 2p-1$,
$0\leq\ell\leq p-1$. We calculate the PBW-basis action on the basis
elements in the projective modules $\PP^{\pm}_s$ (see
\eqref{left-proj-basis-plus} and \eqref{left-proj-basis-min}). Here,
we closely follow to the similar but more complicated calculation
in~\cite{[FGST3]}. We use the well-known identity (see,
e.g.,~\cite{[ChP]})
  \begin{equation}\label{eq:EmFm-prod2}
    F^mE^m = \prod_{i=0}^{m-1}
    \Bigl(\cas-\mfrac{\q^{2i+1}K+\q^{-2i-1}K^{-1}}{(\q-\q^{-1})^2}
    \Bigr),\qquad m<p,
  \end{equation}
where the Casimir element is
\begin{equation*}
\cas= FE+\mfrac{\q K+\q^{-1}K^{-1}}{(\q-\q^{-1})^2}.
\end{equation*}

Using~\eqref{eq:EmFm-prod2}, we calculate the action of $F^m E^m$ for $1\leq m\leq p -1$ 
on $\toppr^{(+,s)}_n$ and $\toppr^{(-,s)}_k$ with the result
\begin{multline}\label{FmEm-toppr+}
F^m E^m \toppr^{(+,s)}_n = \prod_{i=0}^{m-1}
\Bigl(\botpr^{(+,s)}_n \ffrac{\dd}{\dd\toppr^{(+,s)}_n}
+[s+i-n][n-i]\Bigr) \toppr^{(+,s)}_n\\
{}= \Bnm_{n,m}^{+}(s)\botpr^{(+,s)}_n + \Tnm_{n,m}^{+}(s)\toppr^{(+,s)}_n,
\quad 0\leq n\leq s-1,
\end{multline}
and
\begin{multline}\label{FmEm-toppr-}
F^m E^m \toppr^{(-,s)}_k = \prod_{i=0}^{m-1}
\Bigl(\botpr^{(-,s)}_k \ffrac{\dd}{\dd\toppr^{(-,s)}_k}
-[p-s+i-k][k-i]\Bigr) \toppr^{(-,s)}_k \\
{}= \Bnm_{k,m}^{-}(p-s)\botpr^{(-,s)}_k +
\Tnm_{k,m}^{-}(p-s)\toppr^{(-,s)}_k, \quad 0\leq k \leq p-s-1,
\end{multline}
where the coefficients $\Bnm_{n,m}^{\pm}(s)$ and $\Tnm_{n,m}^{\pm}(s)$
are
\begin{itemize}
\item for $1\leq m \leq n$, 
\begin{align}
&\Bnm_{n,m}^{\pm}(s) = (\pm1)^{m-1}
\prod_{j=n-m+1}^{n}[j][s-j]\sum_{i=n-m+1}^{n}\ffrac{1}{[i][s-i]},\label{Bnm-pm}\\
&\Tnm_{n,m}^{\pm}(s)=(\pm1)^{m}\prod_{i=n-m+1}^{n}[i][s-i],\notag
\end{align}
\item for $n+1\leq m \leq p-1$, $\Tnm_{n,m}^{+}(s) = 0,$
\begin{equation}\label{Bnm-plus-2} 
\Bnm_{n,m}^{+}(s) = (-1)^{m-n-1}\qbin{s-1}{n}\qbin{s-1+m-n}{s}([n]![m-n-1]!)^2,
\end{equation}
and, for $k+1\leq m \leq p-1$, $\Tnm_{k,m}^{-}(p-s) = 0$,
\begin{equation}\label{Bnm-min-2}
\Bnm_{k,m}^{-}(p-s) =(-1)^{k} \qbin{s+k}{k}\qbin{s-1}{m-k-1}([k]![m-k-1]!)^2.
\end{equation}
\end{itemize}

\subsection{Idempotents and nilpotents}\label{sec:idem-nilp}
Here, we describe the $\UresSL2$ center in terms of primitive
idempotents and nilpotents. In Thm.~\bref{thm:center-mult}, we use
this basis to calculate the structure constants in the center $\ZZ$
with respect to the Drinfeld and Radford bases.

\begin{prop}[\cite{[FGST]}]\label{prop-center}
  The center $\cZ$ of $\UresSL2$ at $\q=e^{\frac{\rmi\pi}{p}}$ is
  $(3p\,{-}\,1)$-dimensional.  Its associative commutative algebra
  structure is described as follows: there are two ``special''
  primitive idempotents $\idem_0$ and $\idem_p$, \ $p\,{-}\,1$ other
  primitive idempotents $\idem_s$, $1\leq s\leq p-1$, and $2(p\,{-}\,1)$
  elements $\nilp^\pm_s$\ $(1\leq s\leq p\,{-}\,1)$ in the radical such
  that
  \begin{alignat}{2}
    \idem_s\,\idem_{s'}&=\delta_{s,s'}\idem_s,
    &\quad &s,s'=0,\dots,p,\label{idem-idem}\\
    \idem_s\,\nilp^\pm_{s'}&=\delta_{s,s'}\nilp^\pm_{s'},&
    \quad &0\leq s\leq p,~1\leq s'\leq p-1,\label{idem-nilp}\\
    \nilp^\pm_{s}\nilp^\pm_{s'}&=\nilp^\pm_{s}\nilp^\mp_{s'}
    =0,&
    \quad &1\leq s,s'\leq p-1.\label{nilp-nilp}
  \end{alignat}
\end{prop}
We fix the normalization of the nilpotents $\nilp^+_s$
  and $\nilp^-_s$ such that they act as
\begin{equation*}
  \nilp^+_s\,\toppr^{(+,s)}_n=\botpr^{(+,s)}_n,\quad
  \nilp^-_s\,\toppr^{(-,s)}_k=\botpr^{(-,s)}_k
\end{equation*}
in terms of the respective bases in the projective modules $\PP^+_s$
and $\PP^-_{p-s}$ defined in~\bref{module-L} and~\bref{module-P}.

We call $\idem_s$, $\nilp^\pm_s$ the \textit{canonical central
elements}.

\subsubsection{Central elements decomposition}\label{rem:coeffs} 
For any central element $A\in\ZZ$ and its decomposition
\begin{equation}\label{decomp-general}
  A=\sum_{s=0}^{p} a_s \idem_s
  + \sum_{s=1}^{p-1} \bigl(c^+_s \nilp^+_s+c^-_s \nilp^-_s\bigr)
\end{equation}
with respect to the canonical central elements, \textit{the
  coefficient $a_s$ is the eigenvalue of $A$ in the irreducible
  representation $\XX^{+}_s$, the coefficient $c^+_s$ is read off from the relation
  $A\toppr^{(+,s)}_n=c^+_s\botpr^{(+,s)}_n$ in $\PP^+_s$, and
  $c^-_s$, similarly, from the relation
  $A\toppr^{(-,s)}_k=c^-_s\botpr^{(-,s)}_k$ in $\PP^-_{p-s}$}.

\subsection{Calculation of $\Radford(\gamma(s))$}\label{app:Rad-gamma}
The Radford map $\Radford$ on pseudotraces $\gamma(s)$ is given by
\begin{equation*}
\Radford(\gamma(s))=\Radford(s)= \sum_{(\coint)}\Tr_{\PP^{+}_{s}
  \oplus \PP^{-}_{p-s}}(K^{p-1}\coint'\sigma_{s}) \coint'',
\end{equation*}
where the map $\sigma_s$ is defined in~\eqref{sigma-def1}
and~\eqref{sigma-def3}, the projective modules $\PP^{\pm}_{s}$ are
defined in~\bref{proj-mod-base}, and
\begin{multline*}
\Delta(\coint)=\zeta\sum_{r=0}^{p-1}\sum_{s'=0}^{p-1}
\sum_{j=0}^{2p-1} (-1)^{r+s'} \q^{-2(r+1)(s'+1)-r(r+1)-
s'(s'+1)}\times\\ \times F^r E^{p-1-s'} K^{r-p+1+j}\tensor F^{p-1-r}
E^{s'} K^{p-1-s'+j},
\end{multline*}
with $\zeta=\sqrt{\frac{p}{2}}\,\frac{1}{([p-1]!)^2}$. Using
\eqref{FmEm-toppr+} and~\eqref{FmEm-toppr-}, we obtain
\begin{equation}\label{Rad-PBW-new}
\Radford(s) = \bomega^+(s) + \bomega^-(p-s),
\end{equation}
where we introduce
\begin{align}
\bomega^+(s) =&
\sum_{(\coint)}\Tr_{\PP^{+}_{s}}(K^{p-1}\coint'\sigma_{s}) \coint'' =\label{bomega+}\\
&= \alpha_s\zeta \sum_{m=0}^{p-2} \sum_{j=0}^{2p-1}\sum_{n=0}^{s-1} \q^{j(s-1-2n)}
\Bnm^+_{n,p-1-m}(s)F^m E^m K^{j},\notag\\
\bomega^-(p-s) =&
\sum_{(\coint)}\Tr_{\PP^{-}_{p-s}}(K^{p-1}\coint'\sigma_{s}) \coint'' =\label{bomega-}\\
&= \alpha_s \zeta
\sum_{m=0}^{p-2} \sum_{j=0}^{2p-1} \sum_{k=0}^{p-s-1} \q^{j(-s-1-2k)}
\Bnm^-_{k,p-1-m}(p-s) F^m E^m K^{j},\notag
\end{align}
with $\alpha_s$ given in~\eqref{sigma-def3}, and
the coefficients $\Bnm^+_{n,m}(s)$ and $\Bnm^-_{k,m}(p-s)$ are given
in~\eqref{Bnm-pm}, \eqref{Bnm-plus-2} and \eqref{Bnm-min-2}.
\begin{Prop}\label{prop:rad-eval}
The Radford images on the pseudotraces $\gamma(s)$ are decomposed with
respect to the canonical central elements (see~\bref{prop-center}) as
\begin{equation*}
\Radford(s) = a_0 (-1)^{p-s}\ffrac{\sqrt{2p}}{\q^s-\q^{-s}}
\bigl(\idem_s - \ffrac{\q^s+\q^{-s}}{[s]^2}\nilp_s\bigr),
\end{equation*}
where $\nilp_s = \nilp^+_s + \nilp^-_s$.
\end{Prop}
\begin{proof}
In the calculation, we closely follow the strategy proposed
in~\bref{rem:coeffs}. We use \eqref{Rad-PBW-new} to evaluate the action of
$\Radford(s)$ on the modules $\PP^{\pm}_{s'}$, $1\leq s'\leq
p-1$. This action is nonzero only on the module $\PP^{+}_{s}\oplus
\PP^{-}_{p-s}$. Because $\Radford(s)$ is central, it sufficies to
evaluate the action in each direct summand only on any single vector,
which we choose as $\toppr^{(\pm,s)}_0$.
We first evaluate the action of $\bomega^+(s)$
(see~\eqref{Rad-PBW-new} and~\eqref{bomega+}) on $\toppr^{(+,s)}_0$ as
\begin{multline*}
\bomega^+(s) \toppr^{(+,s)}_0
 =\alpha_s\zeta\sum_{n=0}^{s-1} \sum_{j=0}^{2p-1}
 \q^{j(s-1-2n)} \Bnm^+_{n,p-1}(s) K^{j}\toppr^{(+,s)}_0 +\\
{}+ \alpha_s\zeta\sum_{n=0}^{s-1}\sum_{m=1}^{p-2} \sum_{j=0}^{2p-1}
 \q^{j(s-1-2n)} \Bnm^+_{n,p-1-m}(s) K^{j}\prod_{r=1}^{m-1}\bigl(\cas -
 \ffrac{\q^{2r+1}K +
 \q^{-2r-1}K^{-1}}{(\q-\q^{-1})^2}\bigr)\botpr^{(+,s)}_0,
\end{multline*}
where we use \eqref{eq:EmFm-prod2} and the formula $FE\toppr^{(+,s)}_0
= \botpr^{(+,s)}_0$ (see~\bref{module-L}). Then, using~\eqref{Bnm-pm}
and~\eqref{Bnm-plus-2}, we obtain
\begin{multline*}
\bomega^+(s)
\toppr^{(+,s)}_0=\alpha_s(-1)^{p-s-1}\ffrac{p\sqrt{2p}}{[s]^2}\toppr^{(+,s)}_0+\\
{}+ \alpha_s 2p (-1)^p \zeta
 \Bigl(\sum_{m=1}^{s-1}(-1)^{m}\prod_{j=1}^{m}[j][s-j]
 \prod_{r=1}^{p-2-m}[r][s+r] \sum_{i=1}^{m}\ffrac{1}{[i][s-i]}\\
 + (-1)^s\ffrac{[s-1]!}{[s]}\sum_{m=s}^{p-2}[m]![m-s]!
\prod_{r=1}^{p-2-m}[r][s+r]\Bigr) \botpr^{(+,s)}_0,
\end{multline*}
where we set
$\sum_{m=1}^{0}\equiv 0$ and $\prod_{r=1}^{0} \equiv 1$,
and simple calculation gives
\begin{equation}\label{bomega+top+}
\bomega^+(s)\toppr^{(+,s)}_0 
=\alpha_s(-1)^{p-s-1}\ffrac{p\sqrt{2p}}{[s]^2}\toppr^{(+,s)}_0 +
 \alpha_s 2p (-1)^p \zeta (\ff_{s,p} + \bg_{s,p})\botpr^{(+,s)}_0,
\end{equation}
where we introduce the following notation:
\begin{align}
\ff_{s,p}&=(-1)^{s-1}([s-1]![p-s-1]!)^2 \sum_{i=1}^{s-1} \ffrac{1}
{[i][s-i]},\label{ff-def}\\
\bg_{s,p}&=(-1)^s\ffrac{[s-1]![p-s-1]!}{[s]}\sum_{m=1}^{p-s-1}\ffrac{[m+s]![p-s-1-m]!}{[m][s+m]},\notag
\end{align}
and straightforward calculation gives us
\begin{equation*}
\bg_{s,p}=(-1)^{p-1}\ff_{p-s,p}.
\end{equation*}
Therefore, from \eqref{bomega+top+}, we finally obtain
\begin{equation}\label{bomega+top+fin}
\bomega^+(s)\toppr^{(+,s)}_0
=\alpha_s(-1)^{p-s-1}\ffrac{p\sqrt{2p}}{[s]^2}\toppr^{(+,s)}_0
{}+ \alpha_s2p \zeta((-1)^p \ff_{s,p} - \ff_{p-s,p}) \botpr^{(+,s)}_0.
\end{equation}

We next evaluate the action of $\bomega^-(p-s)$
(see~\eqref{Rad-PBW-new} and~\eqref{bomega-}) on $\toppr^{(+,s)}_0$ as
\begin{multline*}
\bomega^-(p-s) \toppr^{(+,s)}_0
 =\alpha_s\zeta\sum_{k=0}^{p-s-1} \sum_{j=0}^{2p-1}
 q^{-2j(k+1)} \Bnm^-_{k,p-1}(p-s)\toppr^{(+,s)}_0\\
{}+ \alpha_s\zeta\sum_{k=0}^{p-s-1}\sum_{m=1}^{p-2} \sum_{j=0}^{2p-1}
 q^{-2j(k+1)} \Bnm^-_{k,p-1-m}(p-s)\prod_{r=1}^{m-1}(\cas -
 \ffrac{q^{2r+1}K + q^{-2r-1}K^{-1}}{(q-q^{-1})^2})\botpr^{(+,s)}_0=0,
\end{multline*}
where we use the simple identity $\sum_{j=0}^{2p-1} q^{-2j(k+1)}=0$.
Hence, using~\eqref{Rad-PBW-new}, \eqref{bomega+top+fin},
\eqref{ff-def}, and the identity
\begin{equation*}
\sum_{i=1}^{p-s-1} \ffrac{1}{[i][s+i]} - \sum_{i=1}^{s-1}
\ffrac{1}{[i][s-i]} = \ffrac{\q^s+\q^{-s}}{[s]^2},
\end{equation*}
we finally obtain the action of $\Radford(s)$ on $\toppr^{(+,s)}_0$ as
\begin{equation*}
\Radford(s)\toppr^{(\pm,s)}_0 = (-1)^{p-s}\ffrac{a_0}{\q-\q^{-1}}
\ffrac{\sqrt{2p}}{[s]}\toppr^{(\pm,s)}_0 +
a_0(-1)^{p-s-1}\ffrac{\q^s+\q^{-s}}{\q-\q^{-1}}
\ffrac{\sqrt{2p}}{[s]^3} \botpr^{(\pm,s)}_0,
\end{equation*}
where we also give the result of analogous calculation of the action
of $\Radford(s)$ on $\toppr^{(-,s)}_0$.
This completes the proof.
\end{proof}

\subsection{The proof of Thm.~\bref{thm:center-mult}} 
\label{sec:proof-Thm}

We recall the definition of the canonical central elements (primitive
idempotents $\idem_s$ and nilpotents $\nilp^{\pm}_s$) given
in~\bref{sec:idem-nilp}.
\begin{prop}[\cite{[FGST]}]
\footnote{We note a misprint in~(\cite{[FGST]}, Lemma 4.5.1):
  $\radmap^-(s)$ should be replaced by $\omega_{p-s}\nilp^-_{p-s}$,
  and $\omega_s$ should be replaced
  by~$(-1)^{p-s-1}\frac{p\sqrt{2p}}{[s]^2}$.}
For $1\leq s\leq p-1$,
 \begin{equation}\label{rad-pm-eval}
    \Radford^{\,+}(s)=\omega_s \nilp^+_s,
    \quad
    \Radford^{\,-}(p-s)=\omega_{s} \nilp^-_{s},
    \qquad
    \omega_s = (-1)^{p-s-1}\ffrac{p\sqrt{2p}}{[s]^2},
  \end{equation}
  \begin{equation}\label{phi-p-idem}
    \Radford^{\,+}(p) = p\sqrt{2p}\,\idem_{p},
    \qquad
    \Radford^{\,-}(p) = (-1)^{p + 1} p \sqrt{2p}\,\idem_{0}.
  \end{equation}
\end{prop}
In Prop.~\bref{prop:rad-eval}, we evaluate $\Radford(s)$ as
\footnote{
We note that $\Radford(s)=a_0 \hrho(s)$ in the terminology of~(\cite{[FGST]}, Sect. 5).}
\begin{equation}\label{rad-eval}
  \Radford(s)=
 a_0 (-1)^{p-s}\ffrac{\sqrt{2p}}{\q^s-\q^{-s}}\bigl(\idem_s -
\ffrac{\q^s+\q^{-s}}{[s]^2}\nilp_s\bigr),
\end{equation}
where $\nilp_s = \nilp^+_s + \nilp^-_s$. 

Using \eqref{rad-pm-eval}-\eqref{rad-eval},
and~\eqref{idem-idem}-\eqref{nilp-nilp}, we calculate the
multiplications \eqref{mult-Rad}-\eqref{mult-Rad-pp} straighforwardly.

The multiplication~\eqref{the-fusion} has been proven
in~\cite{[FGST]}.

\begin{prop}\label{prop:Dr-varphi}
For $1\leq s\leq p-1$,
\begin{equation}\label{Drinf-varphi}
\Drinfeld(s)
= a_0 \ffrac{1}{\sqrt{2p}}
\sum_{j=1}^{p-1}(-1)^{p+s+j}(\q^{sj}-\q^{-sj})\Bigl(
\ffrac{p-j}{p}\,\Radford^+(j) - \ffrac{j}{p}\,\Radford^-(p-j)\Bigr).
\end{equation}
\end{prop}
\begin{proof}
We calculate $\Drinfeld(s)$ using the
$\modS$-transformation~\eqref{mod-center-S}, the identity
$\modS^2|_{\ZZ}=\id$, and~\eqref{rad-eval}. This gives $\Drinfeld(s)$ as
\begin{equation}\label{Dr-modS-Rad}
\Drinfeld(s)=\modS(\Radford(s)),\qquad 1\leq s\leq p-1.
\end{equation}
In the formula~\eqref{rad-eval}, the primitive idempotents $\idem_s$
  and the nilpotents $\nilp_s$ can be linearly expressed
  with respect to the Drinfeld-basis as
\begin{multline*}
\idem_s =
\ffrac{1}{2p^2}\Bigl( \sum_{j=1}^{p-1}(-1)^{j-1}\bigl((p+1-j)(\q^{s(j-1)}+\q^{-s(j-1)})
-(p-1-j)(\q^{s(j+1)}+\q^{-s(j+1)})\bigr)\\
\times \bigl(\Drinfeld^+(j) +
(-1)^{p-s}\Drinfeld^-(j) \bigr)
{}-(\q^{s}+\q^{-s})\bigl((-1)^{p-s}\Drinfeld^+(p)
+ \Drinfeld^-(p)\bigr)\Bigr),
\end{multline*}
and
  \begin{equation*}
    \nilp_s=(-1)^{p-s-1}\ffrac{[s]^2}{2p^2}
    \Bigl(\sum_{j=1}^{p-1}
    (-1)^{p+s+j}\bigl(\q^{js} + \q^{-js}\bigr) \vvarkappa(j)+
    \Drinfeld^+(p) + (-1)^{p-s}\Drinfeld^-(p) \Bigr),
  \end{equation*}
where we introduce the notation
\begin{equation}\label{kappa-def}
\vvarkappa(j)=\Drinfeld^+(j)+\Drinfeld^-(p-j),\qquad 1\leq j\leq p-1.
\end{equation}
We thus obtain
\begin{equation}\label{Rad-rho}
  \Radford(s)= a_0 \ffrac{1}{\sqrt{2p}} \sum_{r=1}^{p-1}(-1)^{r + s +
  p}(\q^{rs}-\q^{-rs})\Bigl(\ffrac{p-r}{p}\,\Drinfeld^{+}(r) -
 \ffrac{r}{p}\,\Drinfeld^{-}(p-r)\Bigr).
\end{equation}
Hence, we get $\Drinfeld(s)$ as in~\eqref{Drinf-varphi} using
\eqref{Dr-modS-Rad} and~\eqref{mod-center-S}.
\end{proof}

Prop.~\bref{prop:Dr-varphi} obviously gives, for $1\leq r,s\leq p-1$,
\begin{equation*}
\Drinfeld(r)\Drinfeld(s)=0,\qquad \Drinfeld^{\pm}(p)\Drinfeld(s)=0.
\end{equation*}

We next calculate the multiplications \eqref{mult-Dr+-NDr}
and~\eqref{mult-Dr--NDr}.
\begin{lemma}\label{lem:NDrinf-gen}
We have, for $1\leq s\leq p-1$,
\begin{gather}
\Drinfeld(s) = \Drinfeld^+(s)\Drinfeld(1),\label{Dr-Dr+1}\\
\Drinfeld(s) = -\Drinfeld^-(p-s)\Drinfeld(1).\label{Dr-Dr-1}
\end{gather}
Therefore, we have
\begin{equation}\label{kappa-drinf}
\vvarkappa(r)\Drinfeld(s)=0,
\end{equation}
with $\vvarkappa(s)$ given in~\eqref{kappa-def}.
\end{lemma}
\begin{proof}
We have, for $1\leq s\leq p$,
\begin{multline}\label{Drinf+-canon}
\Drinfeld^+(s)= s\idem_p + (-1)^{s-1}s\idem_0 +
(-1)^s\sum_{j=1}^{p-1}\Bigl( -\ffrac{[sj]}{[j]}\idem_j\; +\\
+ \ffrac{(s+1)[(s-1)j] - (s-1)[(s+1)j]}{[j]^3}\nilp_j\Bigr),
\end{multline}
and, for $0\leq s\leq p-1$,
\begin{multline}\label{Drinf--canon}
\Drinfeld^-(p-s)= (p-s)\idem_p + (-1)^{s-1}(p-s)\idem_0 +
(-1)^s\sum_{j=1}^{p-1}\Bigl( \ffrac{[sj]}{[j]}\idem_j\; +\\
+ \ffrac{(p-s-1)[(s-1)j] - (p-s+1)[(s+1)j]}{[j]^3}\nilp_j\Bigr),
\end{multline}
with $\nilp_j = \nilp^+_j + \nilp^-_j$.  Hence, we obtain
\eqref{Dr-Dr+1} and~\eqref{Dr-Dr-1} by use of~\eqref{Drinf-varphi},
\eqref{Drinf+-canon}, and~\eqref{Drinf--canon}.
\end{proof}

To prove \eqref{mult-Dr+-NDr} and~\eqref{mult-Dr--NDr},
we use \eqref{the-fusion} and Lem.~\bref{lem:NDrinf-gen}.
This completes the proof of Thm.~\bref{thm:center-mult}.

\end{document}